\documentclass[prodmode,acmtois]{acmsmall} 

\acmVolume{0}
\acmNumber{0}
\acmArticle{00}
\acmYear{2014}
\acmMonth{0}

\usepackage{grffile}
\usepackage{epstopdf}
\usepackage{amssymb}
\usepackage{amsfonts}
\usepackage{subfigure}

\usepackage{algorithm}
\usepackage{algpseudocode}
\usepackage{graphicx}
\usepackage{graphics}
\usepackage{multirow}
\usepackage{comment}
\usepackage{array}

\begin{document}

\markboth{Seung-Hoon Na}{Two-Stage Document Length Normalization}

\title{ Two-Stage Document Length Normalization for Information Retrieval}
\author{SEUNG-HOON NA
\affil{Busan University of Foreign Studies, South Korea}}


\begin{abstract}
The standard approach for term frequency normalization is based only
on the document length. However, it does not distinguish the verbosity from
the scope, these being the two main factors determining the document
length. Because the verbosity and scope have largely different
effects on the increase in term frequency, the standard approach
can easily suffer from insufficient or excessive penalization depending on
the specific type of long document. To overcome these problems,
this paper proposes two-stage normalization by performing verbosity
and scope normalization separately, and by employing different penalization
functions. In verbosity normalization, each document is pre-normalized
by dividing the term frequency by the verbosity of the document. In
scope normalization, an existing retrieval model is applied in a
straightforward manner to the pre-normalized document, finally leading
us to formulate our proposed verbosity normalized (VN) retrieval
model. Experimental results carried out on standard TREC collections
demonstrate that the VN model leads to marginal but statistically significant improvements over
standard retrieval models.
\end{abstract}





\acmformat{Seung-Hoon Na, 2014. Two-stage document length normalization for information retrieval.}

\begin{bottomstuff}
Author e-mail: nash@bufs.ac.kr\\
This work was partly supported by the IT R\&D program of MSIP/KEIT. [10041807, Development of Original Software Technology for Automatic Speech Translation with Performance 90\% for Tour/International Event focused on Multilingual Expansibility and based on Knowledge Learning] and by the research grant of the Busan University of Foreign Studies in 2014.
\end{bottomstuff}

\def\mathbi#1{\textbf{\em #1}}

\maketitle

\section{INTRODUCTION}

In information retrieval (IR), term frequency is a fundamental and
important component of a ranking model. Intuitively, the larger the
term frequency of a query word in a document, the more likely the document is
to be about the query topic, and thus, the document should have a higher
relevance score. In practice, however, documents are of various
lengths, and the simple approach of preferring documents
with higher term frequency could easily result in an excessive preference
for long documents. To use the term frequency in a fairer approach,
\emph{normalization} of the term frequency has been extensively
investigated by researchers.


With regard to the normalization problem, Robertson and Walker
introduced the verbosity and the scope hypotheses, which state that
document length is mainly determined by two factors --
\emph{verbosity} and \emph{scope} -- as follows
\cite{robertson94,robertson09}:

1) \textbf{Verbosity hypothesis}: ``\emph{Some authors are simply more
   verbose, using more words to say the same thing}
   \cite{robertson09}.''

2) \textbf{Scope hypothesis}: ``\emph{Some authors have more to say:
   they may write a single document containing or covering more ground}
   \cite{robertson09}.''

In this paper, we focus on the difference between the effect of
the verbosity and the scope on the term frequency of a single word. Verbosity,
as the name implies, is related to the burstiness of term frequency,
which helps an already mentioned word in a document get a higher
frequency. Even if a word has a low term frequency in normal
verbosity, its term frequency could increase significantly when the
document has high verbosity. On the other hand, scope mostly involves
the creation of a new word, rather than boosting the term
frequency. Broadening the scope of a document would help unseen words
in a normal document get non-zero frequencies. However, these
non-zero frequencies might not be high. Therefore,
verbosity leads to a significant increase in term frequency,
whereas scope leads to a rather limited increase in term
frequency. In other words, the scope of a document only helps the
occurrence of a new word, and the term frequency of the word is mostly governed by
the verbosity of the document.

Despite this difference between verbosity and scope, standard
normalization is a length-driven approach, i.e., it is based only on the
document length, without distinguishing between verbosity and
scope. As a result, it may suffer from \emph{insufficient} penalization
of a \emph{verbose} document whose length is increased mainly by high
verbosity, and \emph{excessive} penalization of a \emph{broad}
document whose length is mainly derived from the broad scope.

In the light of this addressed difference, this paper argues that verbosity
and scope should be normalized separately by employing different penalization functions.
To achieve this, we propose a two-stage normalization approach. We first perform
\emph{verbosity normalization} for each document by linearly dividing
the term frequency by the verbosity, thus obtaining a \emph{verbosity-normalized
  document representation}. We then perform \emph{scope normalization}, in which an
existing retrieval model is applied to this
verbosity-normalized document representation. The final model obtained is called
a \emph{verbosity-normalized} (VN) \emph{retrieval model}.


Furthermore, we examine whether the proposed VN
retrieval model resulting from two-stage normalization performs
the desired separate penalizations. Toward this end, we first select three
popular retrieval models -- the Okapi model \cite{robertson95}, the
Dirichlet-prior (DP) smoothed language model \cite{zhai01}, and the Markov random field (MRF) model \cite{metzler05mrf}
-- and then perform \emph{comparative axiomatic analysis} of the original and the VN
retrieval models, under the setting of the axiomatic framework
introduced in \cite{fang04,fang11}. The analysis results
confirm that the VN model indeed performs the desired separate
normalizations, i.e., a \emph{strict} penalization of verbosity-increased
documents and a \emph{relaxed} penalization of scope-broadened documents.

The results of experiments carried out on standard TREC test collections
show that the VN retrieval models are significantly better than the
original models. The experimental results support our motivating
argument that the verbosity and scope should be handled separately using
different penalization functions.

The remainder of this paper is organized as follows. Section \ref{section_related_work} describes
previous studies. Section \ref{section_two_stage_normalization} describes the proposed two-stage
normalization approach and presents the VN retrieval models for DP, Okapi,
and MRF. Section \ref{section_comparative_axiomatic_analysis} presents the main results of the analysis of retrieval
models under standard length normalization constraints. Sections \ref{section_experiments}, \ref{section_experimental_results}, \ref{section_application_lower_bounding_term_freq_normalization} present the experimental setting and results. Sections \ref{section_conclusion} concludes.

\section{PREVIOUS WORK}
\label{section_related_work}


\citeN{singhal96} recognized that simply dividing the term frequency by
the document length leads to the over-penalization problem in long documents.
To overcome this problem, they proposed
\emph{pivoted normalization}, in which a pivoted length is used to normalize the term frequency by adding a constant pivot factor (i.e., average document length) to the original document length. Pivoted normalization had originally been introduced in Okapi's model \cite{robertson95}, before it was formalized and named by \citeN{singhal96}. Because pivoted normalization yields successful results, it has been explicitly adopted by other retrieval models, such as the INQUERY system \cite{callan92,inquery00}. A similar relaxed type of normalization has been commonly used in more recent retrieval models -- normalization 2 in the divergence from randomness (DFR) retrieval framework \cite{amati02} and the smoothed document length in DP \cite{zhai01}.

\citeN{fang04} formally and mathematically defined IR heuristics, drawn from ranking characteristics most commonly used by existing retrieval models, thereby proposing a novel direction for an
axiomatic approach to IR. The retrieval heuristics
defined in the axiomatic approach have been used to define a new
retrieval model inductively \cite{fang05,clinchant10} and to restrict
the search space for automatically learning a retrieval
function \cite{cummins06}. In addition to original constraints, some studies have explored new constraints
including: semantic term matching constraints \cite{fang06}, the proximity-based matching constraint \cite{tao07},
the burstness-based normalization constraint \cite{clinchant10},
the document frequency constraint for pseudo-relevance feedback \cite{clinchant11},
the feedback term weight constraints for pseudo-relevance feedback \cite{clinchant13},
and the translation probability constraints for translation language models \cite{karimzadehgan12}.
With regard to the length normalization problem, \citeN{fang04} defined three length normalization constraints
(referred to as LNC1, LNC2, and TF-LNC), demonstrating analytically that popular retrieval models satisfy all
these normalization constraints at least for content-bearing words.


Our argument that different normalization functions should be used for
verbosity and scope was also proposed by
\cite{robertson94,robertson09}, in a more restricted manner, as
follows: ``\emph{The verbose hypothesis suggests that we should simply
  normalize term frequencies by dividing by document length, while the
  scope hypothesis, on the other hand, suggests the opposite}
\cite{robertson09}.'' That is, they suggest that a retrieval function
does not necessarily penalize a long document when it has a
broad scope. A similar argument was also made by \cite{na08}. Our
suggestion, however, is that we still need the
penalization for scope, but in a much more relaxed manner. In this
sense, our argument can be regarded as a generalization of the
previous arguments.

To the best of our knowledge, one of the first approaches for two-stage
normalization is \emph{pivoted unique normalization}, suggested by
\cite{singhal96}. In their approach, the term frequency is first
normalized on the basis of a nonlinear function by using the average term
frequency (which corresponds to verbosity normalization), and the
normalized term frequency is then further divided by a pivoted unique
length (which corresponds to scope normalization). However, it remains
unclear how their approach can be generalized to other retrieval
models.


Going beyond the aforementioned existing works, we propose a
generalized two-stage normalization approach, arguing more clearly
that the term frequency should be penalized differently, depending on
whether a document is long because of the verbosity or the scope. Our approach is
not limited to a specific retrieval model or a specific measure of
the verbosity or scope. We also analytically present the
retrieval heuristics realized by two-stage normalization, by performing a comparative axiomatic
analysis under the setting of standard normalization constraints
suggested by \citeN{fang04},

\sloppypar{ It is noteworthy that the Okapi and the DFR retrieval framework \cite{amati02} can be
  considered as another type of two-stage normalization. According to
  the derivation by \cite{he03}, the first step normalizes the term
  frequency by a relaxed document length using $tfn$ =
  $c(w,d)/\left(k_1 \left((1-b) + b \cdot |d|/avgl\right)\right)$ in
  Okapi and $tfn$ = $c(w,d) log \left(1+c \cdot avgl/|d| \right)$ in
  DFR, and the second step further normalizes $tfn$ by
  $tfn/(tfn+1)$. The first step uses the document length, thereby
  performing a mixed normalization of verbosity and scope, and the
  second step roughly performs verbosity normalization by
  preventing a document with high $tfn$ from getting a very large
  score. However, this is not the case in our approach, which further
  distinguishes between the verbosity and the scope.  }


Interestingly, passage retrieval can also be viewed as two-stage
normalization \cite{salton91,salton93,callan94,allan95,mittendorf94,kaszkiel97,kaszkiel99,liu02,bendersky08,na08b,bendersky10,lv09,lv10,krikon10,krikon11}.
Because scopes are more similar in passages themselves than in documents, using
passages itself can be considered as a type of scope
normalization. Thereafter, applying an existing retrieval method to score
each passage corresponds to verbosity normalization.

Recently, \citeN{lv11a} and \citeN{lv11b} observed that when documents are extremely long, the score gap calculated as the difference between scores when a query term is present and when it is absent in a document, could be infinitely close to zero or negative. As a result, extremely long documents tend to be overly penalized. To ensure a desirable score gap between documents that match and do not match a query term, \citeN{lv11b} proposed \emph{lower-bounding} term frequency normalization, which can be described as follows: (1) A \emph{pseudo score gap} between documents that match and do not match a query term is newly introduced as a document-independent factor. (2) For each query term, the pseudo score gap is added to the original document score only when the document matches the query term, whereas the original document score is left unchanged for a document that does not match the query term\footnote{The same scoring function can be equivalently implemented by redefining a within-document scoring function for both cases (i.e., either a document matches a query or it does not), as formulated in \cite{lv11b}.}. Importantly, \citeN{lv11b} closely examined the underlying principles of their proposed normalization, after which they proposed the constraints \textbf{LB1} and \textbf{LB2} as extensions of the existing formal heuristics used in \cite{fang11}. According to their axiomatic analysis, all modified retrieval functions proposed in \cite{lv11b} unconditionally or more easily satisfy the lower bounds (LBs) without violating the original constraints of \cite{fang11}, whereas existing functions do not satisfy the LBs. Experiment results showed that all modified retrieval functions showed statistically significant improvements, especially for verbose queries. In contrast to our work, the lower-bounding normalization proposed in \cite{lv11b} uses only the document length. However, in our case, we distinguish the verbosity of the document from the scope. In addition, the new constraints used in \cite{lv11b} are complementary to the existing length normalization constraints (LNCs), whereas our work emphasizes the need to pursue a new generation of LNCs.

\subsection{Novel Contributions beyond Our Prior Work}
\label{section_related_work_on_prior_work}

In \cite{na08}, the initial form of the two-stage normalization approach was presented to modify language modeling approaches by introducing the \emph{pseudo document model}. However, \cite{na08} were not aware of the importance of the pseudo document model as a generalized solution for handling the addressed problem. In addition, \cite{na08} suggested a rather harsh retrieval constraint called TNC, which is too strong to be satisfied by even their own proposed method. Given the previous presentation of \cite{na08}, it thus remained unclear how the presented normalization yields some of the reported improved performances, and how it can be generalized to other retrieval models. Building on our prior work, novel contributions of this paper are listed in the following:

\begin{itemize}
  \item \emph{Generalized} two-stage normalization (Section \ref{section_two_stage_normalization}), which was not explicitly argued and not fully formalized in \cite{na08}. With the explicit formulation, we now correctly understand VN-DP as a specific instance of two stage normalization.
  \item Extensions to other models -- Okapi and MRF (Section \ref{section_two_stage_normalization}), as a result of the proposed generalized normalization
  \item Analytically capturing retrieval heuristics of two-stage normalization by performing \emph{comparative axiomatic analysis} (Section \ref{section_comparative_axiomatic_analysis} \& Appendices C, D, and E) under the standard constraint setting of \cite{fang04,fang11}
  \item \emph{LengthPower} as a novel scope measure (Section \ref{section_two_stage_normalization}). Using LengthPower, we have an unified view of language modeling approaches by considering both JM and DP as special cases of VN-DP.
  \item \emph{Comparison with lower bounding term frequency normalization} (Section \ref{section_application_lower_bounding_term_freq_normalization})
\end{itemize}



\section{TWO-STAGE NORMALIZATION}
\label{section_two_stage_normalization}

In this section, we describe our proposed two-stage normalization in
detail, and apply it to the DP, Okapi, and MRF approaches, as case studies.

\subsection{Verbosity Normalization}
The following are notations commonly used in this paper. \\
$\bullet$ $V$: ${w_1,w_2, \cdots,w_{|V|}}$, Set of all words \\
$\bullet$ $N$: Number of documents in a given collection \\
$\bullet$ $C$: A given collection, consisting of $d_1, \cdots, d_N$. Often, we also use $C$ to refer to the concatenated representations of all documents in $C$. \\
$\bullet$ $df(w)$: Document frequency of $w$ \\
$\bullet$ $d$ (or $q$): A given document (or a query) \\
$\bullet$ $c(w,d)$ (or $c(w,q)$): Term frequency of word $w$ in document $d$ (or query $q$) \\
$\bullet$ $c(w,C)$: Term frequency of word $w$ in collection $C$ defined by $\sum_{d \in C} c(w,d)$\\
$\bullet$ $idf(w)$: Term discrimination value of $w$ such as IDF \\
$\bullet$ $|d|$: Length of document $d$, defined by $\sum_{w \in V} c(w,d)$ \\
$\bullet$ $|C|$: Length of collection $C$, defined by $\sum_{w \in V} c(w,C)$ (for brevity of notation, $C$ is either the set of documents or the concatenated representation of documents, depending on context) \\
$\bullet$ $s(d)$: Scope of document $d$ ($ s(d) \leq|d|$ )\\
$\bullet$ $v(d)$: Verbosity of document $d$ \\
$\bullet$ $avgl,avgv,avgs$: Average length, verbosity, and scope, respectively, of documents in the collection.


Motivated by the verbosity and the scope hypotheses, we first assume that the document length is decomposed into the verbosity and the scope, thereby providing the following simplified formula:
\begin{equation}
|d| = v(d) s(d)
\label{eq_def_length_in_terms_verbosity_scope}
\end{equation}
As a result, we can formulate $v(d)$ in terms of $s(d)$ and $|d|$ as follows:
\begin{equation}
v(d)  = \frac{|d|}{s(d)}
\label{eq_def_verbosity}
\end{equation}
The derivation of Eq. (\ref{eq_def_verbosity}) is presented in Appendix A.

In verbosity normalization, the original term frequency is normalized by dividing it by the verbosity of the document. To formally describe verbosity normalization, let $\phi$ be a \emph{verbosity normalization operator}; $\phi(d)$, the \emph{verbosity-normalized document representation} of $d$\footnote{In our notation, the verbosity normalization operator is applied not to document itself but instead to the document representation. In this paper, the document representation is assumed to be a vector of its term frequencies (and either bigram or proximal term frequencies). For general purposes, the verbosity normalized operator needs to be extended such that it can be applied to advanced document representation such as a sequence of words, so that it can be useful for the proximity-based or location-based search.}, which is the document transformed by applying the operator $\phi$ to all words in a document $d$; and $c(w, \phi(d))$, the \emph{verbosity-normalized term frequency} of word $w$. Then, verbosity normalization refers to the process of obtaining $c(w, \phi(d))$ for word $w$, using the following formula:
\begin{equation}
c(w,\phi(d))  = k \frac{c(w,d)}{v(d)}
\label{eq_def_verbosity_normalized_TF}
\end{equation}
where $k$ is a verbosity scaling parameter. By substituting Eq. (\ref{eq_def_verbosity}) into Eq. (\ref{eq_def_verbosity_normalized_TF}), $c(w,\phi(d))$ becomes
\begin{displaymath}
c(w,\phi(d))  = k \frac{c(w,d) \cdot s(d) }{|d|}
\label{eq_def_verbosity_normalized_TF2}
\end{displaymath}
The resulting normalized term frequency is not only inversely proportional to the document length but is also proportional to the scope of the document.


\subsection{Scope Normalization}
\label{section_scope_normalization}

For scope normalization, 
we need to consider a more relaxed function than that for verbosity normalization.
We first note that the scope of an original document is the verbosity-normalized length of the document, as follows:
 \begin{displaymath}
|\phi(d)| = \sum_{w \in V} c(w,\phi(d)) = k \frac{\sum_{w \in V} c(w,d) \cdot s(d)}{|d|} = k \cdot s(d)
\label{eq_length_VN_document}
\end{displaymath}
Furthermore, existing retrieval models perform a type of relaxed normalization by using their pivoted length or smoothed length.
Thus, instead of developing a new function, we perform scope normalization by straightforwardly applying an existing retrieval model to the verbosity-normalized document representation $\phi(d)$.  Formally, let $f(d,q)$ be the original retrieval function that gives a score to $d$, for query $q$. Applying two-stage normalization to $f(d,q)$ gives $f(\phi(d),q)$, which is obtained by replacing $c(w,d)$ used in all terms in $f(d,q)$ with $c(w,\phi(d))$ for all documents in the collection. We call $f(\phi(d),q)$ a \textbf{VN} (verbosity-normalized) \textbf{retrieval model} or a \textbf{VN scoring function}.  

\subsection{Examples of Verbosity-Normalized Retrieval Models}
In this section, we present the application of two-stage normalization to the DP, Okapi, and MRF approaches.

\subsubsection{Dirichlet-prior (DP)}
DP performs Bayesian smoothing on a multinomial language model \cite{zhai01}, for which the conjugate prior is the Dirichlet distribution with the following parameters:
\begin{equation}
\left( \mu p(w_1|C), \mu p(w_2|C), \cdots, \mu p(w_{|V|}|C) \right)
\label{eq_Dirichlet_prior}
\end{equation}
The Bayesian priors using the parameters of Eq. (\ref{eq_Dirichlet_prior}) give the following smoothed model of document $d$:
\begin{displaymath}
P(w|\phi(d)) = \frac{c(w,d)+\mu p(w|C)}{|d| + \mu}
\label{eq_Dirichlet_smoothed_model}
\end{displaymath}
and the following scoring function for a given query $q$ \cite{zhai01}:
\begin{eqnarray}
\sum_{w \in q \cap d} ln \left( 1 + \frac{c(w,d)}{\mu \cdot p(w|C)} \right) + |q| \cdot ln\left( \frac{\mu}{|d|+\mu}\right) \nonumber
\label{eq_score_Dirichlet_smoothed_model}
\end{eqnarray}
The VN model $f(\phi(d),q)$ is assumed to employ the following \emph{document-specific} conjugate prior:
\begin{equation}
\left( \mu v(d) p(w_1|C), \mu v(d) p(w_2|C), \cdots, \mu v(d) p(w_{|V|}|C) \right)
\label{eq_DirichletVN_prior}
\end{equation}
In other words, the more verbose $d$ is, the larger is the prior probability used. 
A detailed justification for Eq. (\ref{eq_DirichletVN_prior}) is presented in Appendix B.
These modified Bayesian priors using the parameters of Eq. (\ref{eq_DirichletVN_prior}) give the following smoothed model:
\begin{equation}
P(w|d) = \frac{c(w,d)+\mu v(d) p(w|C)}{|d| + \mu v(d)}
\label{eq_DirichletVN_smoothed_model}
\end{equation}
We simply use $k = 1$ in Eq. (\ref{eq_def_verbosity_normalized_TF}), because the scaling parameter $k$ of $c(w,\phi(d))$ is absorbed into the smoothing parameter $\mu$.
Then, Eq. (\ref{eq_DirichletVN_smoothed_model}) becomes
\begin{eqnarray}
P(w|\phi(d)) & = &\frac{c(w,\phi (d)) +\mu \cdot p(w|C)}{|\phi(d)| + \mu}
\label{eq_DirichletVN_smoothed_model_using_operator}
\end{eqnarray}
Eq. (\ref{eq_DirichletVN_smoothed_model_using_operator}) is the same as the equation obtained by replacing $c(w,d)$ with $c(w,\phi(d))$.

Using Eq. (\ref{eq_DirichletVN_smoothed_model}), the resulting retrieval function is given as
\begin{eqnarray}
 \sum_{w \in q \cap d} c(w,q) ln\left( 1 + \frac{c(w,d)}{\mu \cdot p(w|C)} \frac{s(d)}{|d|}\right) + |q| \cdot ln\left( \frac{\mu}{s(d)+\mu}\right)  \nonumber
\end{eqnarray}
which is called \textbf{VN-DP}\footnote{The formula of VN-DP is equivalent to the modified Dirichlet-prior smoothing suggested by \cite{na08}.
}.

\subsubsection{Okapi}
Okapi's BM25 retrieval formula, as presented by \cite{robertson95}, is
\begin{displaymath}
\sum_{w \in q \cap d} \left\{ \frac{(k_3+1) c(w,q)}{k_3 + c(w,q)} ln\left(\frac{N - df(w) + 0.5}{df(w) + 0.5}\right) tf_{BM25}(w,d) \right\}
\end{displaymath}
where the term frequency component $tf_{BM25}(w,d)$ is
\begin{displaymath}
tf_{BM25}(w,d) = \frac{(k_1+1) c(w,d)}{k_1 \left( (1-b) + b \frac{|d|}{avgl}\right) + c(w,d)}
\label{eq_Okapi_term_frequency}
\end{displaymath}
Here, $k_1$, $k_3$, and $b$ are constants. In the VN model, the IDF part is not changed; however,  $tf_{BM25}(w,d)$ is modified to $tf_{BM25}(w,\phi(d))$ obtained by replacing $c(w,d)$ with $c(w,\phi(d))$, as follows:
\begin{displaymath}
tf_{BM25}(w,\phi(d)) = \frac{(k_1+1) c(w,\phi(d))}{k_1 \left( (1-b) + b \frac{|\phi(d)|}{avgs}\right) + c(w,\phi(d))}
\label{eq_OkapiVN_term_frequency}
\end{displaymath}
As in the case of DP, we assume the scale parameter $k$ to be 1, because it is absorbed into $k_1$, resulting in the following final form:
\begin{eqnarray}
tf_{BM25}(w,\phi(d)) &= & \frac{(k_1+1) c(w,d)}{k_1 |d| \left( (1-b) \frac{1}{s(d)}+ b \frac{1}{avgs}\right) + c(w,d)} \nonumber
\label{eq_OkapiVN_term_frequency_derivation}
\end{eqnarray}
The modified Okapi function by using $tf_{BM25}(w,\phi(d))$ for $tf_{BM25}(w,d)$ is called \textbf{VN-Okapi}.

\subsubsection{Markov Random Field (MRF)}

MRFs are undirected graphical models that are used to define joint distributions over a set of random variables. The use of MRFs for IR was suggested by \cite{metzler05mrf,metzler07b}, going beyond the simplistic bag of words assumption, by explicitly modeling the term dependency among query words. Thus far, three different variants of the MRF model have been suggested according to the type of dependency assumed among query words -- full independence, sequence dependence, and full dependence. This paper focuses on \emph{sequence dependence}, which has been widely used in many recent works
\cite{metzler07,lease09,bendersky10mrf,wang10,lang10,bendersky11}, because of its good balance between effectiveness and efficiency.

To formally present the ranking function of the sequential dependence, suppose that $q$ is a sequence of $m$ terms $q_1 \cdots q_m$. According to the original framework, the relevance score of a document $d$ is given by \cite{metzler05mrf}
\begin{equation}
f(d,q) = \lambda_T \sum_{q_i \in q} f_T(d,q_i) + \lambda_O \sum_{q_i q_{i+1} \in q} f_O (d,q_i q_{i+1})
+ \lambda_U \sum_{q_i q_{i+1} \in q} f_U(d,q_i q_{i+1})
\label{eq_score_MRF_model}
\end{equation}
where we have the constraint $\lambda_T + \lambda_O + \lambda_U$ = 1, and $f_T(d,q_i)$, $f_O(d,q_i q_{i+1})$ and $f_U(d,q_i q_{i+1})$ are called the \emph{feature functions} of the term, \emph{ordered phrase}, and \emph{unordered phrases}, respectively. Table \ref{tbl_MRF_feature_fuction} presents the definition of each feature function \cite{metzler05mrf}.

\begin{table}
\tbl{Feature functions used in the MRF model.
$c_{\#1}(q_i q_{i+1},d)$ indicates the number of times that the \emph{exact phrase} $q_i q_{i+1}$ occurs in document $d$, and $c_{\#un8}(q_i q_{i+1},d)$ indicates the number of times that both terms $q_i$ and $q_{i+1}$ appear \emph{ordered} or \emph{unordered} within a window with a span of 8.\label{tbl_MRF_feature_fuction}}{%
\begin{tabular}{||c|c||}\hline
Feature & Value \\\hline
$f_T(d,q_i)$ & $ \ln \left[ \frac{c(q_i,d) + \mu_T \frac{c(q_i,C)}{|C|}}{|d| + \mu_T} \right]$ \\\hline
$f_O(d,q_i q_{i+1})$ & $ \ln \left[ \frac{c_{\#1}(q_i q_{i+1},d) + \mu_O \frac{c_{\#1}(q_i q_{i+1},C)}{|C|}}{|d| + \mu_O} \right]$ \\\hline
$f_U(d,q_i q_{i+1})$ & $ \ln \left[ \frac{c_{\#un8}(q_i q_{i+1},d) + \mu_U \frac{c_{\#un8}(q_i q_{i+1},C)}{|C|}}{|d| + \mu_U} \right]$ \\\hline
\end{tabular}
}
\end{table}

Following the original framework \cite{metzler05mrf}, we assume that $\mu_T$, $\mu_O$, and $\mu_U$ are the same, i.e., $\mu_T = \mu_O = \mu_U = \mu$, unless otherwise stated. We refer to the retrieval function in Eq. (\ref{eq_score_MRF_model}) as \textbf{MRF}.

To derive a VN retrieval model $f(\phi(d),q)$ for MRF, we replace the original term frequencies with the verbosity normalized ones. For this purpose, let $c_{\#1}(q_i q_{i+1},\phi(d))$ and $c_{\#un8}(q_i q_{i+1},\phi(d))$ be \emph{VN ordered} and \emph{unordered phrase term frequencies} for $q_i q_{i+1}$, respectively. Similar to the definition of VN term frequency in Eq. (\ref{eq_def_verbosity_normalized_TF}), these VN phrase term frequencies are defined as follows:
\begin{equation}
c_{\#1}(q_i q_{i+1},\phi(d)) = k \frac{c_{\#1}(q_i q_{i+1},d)}{v(d)}
\label{eq_VN_ordered_phrase_term_freq}
\end{equation}

\begin{equation}
c_{\#un8}(q_i q_{i+1},\phi(d)) = k \frac{c_{\#un8}(q_i q_{i+1},d)}{v(d)}
\label{eq_VN_unordered_phrase_term_freq}
\end{equation}

Furthermore, let $f_T( \phi(d),q_i)$, $f_O( \phi(d),q_i q_{i+1})$, and $f_U( \phi(d), q_i q_{i+1})$ be \emph{VN feature functions} that correspond to original feature functions. Table \ref{tbl_VN_MRF_feature_fuction} describes the definition of each VN feature function.
\begin{table}
\tbl{Verbosity-normalized feature functions used in the VN-MRF model.\label{tbl_VN_MRF_feature_fuction}}{%
\begin{tabular}{||c|c||}\hline
Feature & Value \\\hline
$f_T(\phi(d),q_i)$ & $ \ln \left[ \frac{\frac{c(q_i,d)}{v(d)} + \mu_T \frac{c(q_i,C)}{|C|}}{s(d) + \mu_T} \right]$ \\\hline
$f_O(\phi(d),q_i q_{i+1})$ & $ \ln \left[ \frac{\frac{c_{\#1}(q_i q_{i+1},d)}{v(d)} + \mu_O \frac{c_{\#1}(q_i q_{i+1},C)}{|C|}}{s(d) + \mu_O} \right]$ \\\hline
$f_U(\phi(d),q_i q_{i+1})$ & $ \ln \left[ \frac{\frac{c_{\#un8}(q_i q_{i+1},d)}{v(d)} + \mu_U \frac{c_{\#un8}(q_i q_{i+1},C)}{|C|}}{s(d) + \mu_U} \right]$ \\\hline
\end{tabular}
}
\end{table}
As in the case of VN-DP, $k$ is assumed to be 1 in all VN feature functions, because it is absorbed to $\mu_T$, $\mu_O$, or $\mu_U$. Finally, we obtain the scoring function for the VN model $f(\phi(d),q)$ of MRF as follows:
\begin{equation}
f(\phi(d),q) =
\lambda_T \sum_{q_i \in q} f_T(\phi(d),q_i) + \lambda_O \sum_{q_i q_{i+1} \in q} f_O (\phi(d),q_i q_{i+1})
+ \lambda_U \sum_{q_i q_{i+1} \in q} f_U( \phi(d),q_i q_{i+1})
\label{eq_score_VN_MRF_model}
\end{equation}
The MRF model using Eq. (\ref{eq_score_VN_MRF_model}) is referred to as \textbf{VN-MRF}.

\subsection{Scope Measure}
The remaining problem is how to compute the scope of a document $s(d)$. In this study, we adopt three different approaches -- length power, the number of unique terms, and entropy power.

\subsubsection{LengthPower}

As mentioned in the introduction, according to the scope hypothesis,
the document length is affected by the scope: the broader the scope of a
document, the longer the document is, when its verbosity is assumed to be fixed. Therefore, the document length could
possibly be used as a scope measure according to the scope
hypothesis. To derive such a length-based measure, suppose that the
scope of a document is a function of document length, i.e., $s(d) =
g(|d|)$. Many variants exist for such a function; however, the verbosity
and the scope hypotheses help us restrict the possible space for $g(|d|)$,
given the following two necessary constraints:

$\bullet$ \textbf{SC1}: Scope $g(|d|)$ is a non-decreasing function of $|d|$. \\

$\bullet$ \textbf{SC2}: Verbosity $|d|/g(|d|)$ is a non-decreasing function of $|d|$. \\

To obtain such a scope measure that would satisfy both SC1 and SC2, we use Heap's law, which is given as follows \cite{heaps78}\footnote{The Heaps law predicts the number of unique terms in a document from the document length, i.e., the number of unique terms in a corpus increases according to a $k \cdot |d|^\beta$ relationship to the document length. Because the number of unique terms can be used as a scope measure to indicate how broad the topic of the document is, as presented in Section \ref{subsec_uniqlength}, we use the formula of the Heaps law to approximately predict the number of unique terms using only the document length.}:
\begin{displaymath}
l_\beta(d) = |d|^\beta
\label{eq_Heap_scope}
\end{displaymath}
where $\beta$ is an additional constant\footnote{The original form of Heap's law is $\kappa |d|^\beta$, containing the additional parameter $\kappa$. Here, we assume that $\kappa$ is absorbed in $k$.}.

The possible range of $\beta$ is $0\leq \beta \leq 1$, from SC1 and SC2. Otherwise, $s(d)$ (or $v(d)$) violates SC1 (or SC2) if $\beta < 0$ (or $\beta > 1$). This length-based scope measure $l_\beta(d)$ exactly degenerates into the original unnormalized representation, as a special case, when $\beta$ = 1 and $k$ = 1, in which case $l_\beta(d)$ = $|d|$,  $v(d)$ = 1, and $s(d) = |d|$.
The scope measure using $l_\beta(d)$ is called \textbf{LengthPower} in this paper.

\subsubsection{UniqLength }
\label{subsec_uniqlength}
Another useful scope measure is the number of unique terms $u(d)$, defined as $|\left\{w|w \in d\right\}|$. This is reasonable, because a different topic is described using a domain-specific vocabulary or named entities. The more unique terms used in a document, the larger is the scope of the document. The scope measure $u(d)$ is referred to as \textbf{UniqLength} in this paper.

\subsubsection{EntropyPower }
The third scope measure is an entropy-based metric. Previously, the entropy of a document was used to define the homogeneous measure of a document \cite{bendersky08}, which corresponds to the opposite concept of scope. Another entropy-based metric is the entropy power defined by the exponential of the entropy, which was initially exploited in \cite{kurland05} to construct the document structure. We compared the entropy with the entropy power in our preliminary experiments and found that the latter outperformed the former because of its similarity to document length or the number of unique terms. Thus, we choose entropy power as our entropy-based metric, and it is defined as follows:
\begin{displaymath}
h(d) = \left\{\begin{array}{cc} \exp \left( -\sum_w p_{ml}(w|d) ln\left(p_{ml}(w|d)\right)\right)  & if |d| \geq 1 \\
        0   & otherwise \end{array}\right.
\end{displaymath}
where $p_{ml}(w|d)$ is defined by $c(w,d)/|d|$, which is the maximum likelihood estimation (MLE) of the document language model for $d$.  The scope measure $h(d)$ is called \textbf{EntropyPower} in this paper.


\section{RETRIEVAL HEURISTICS of VN RETRIEVAL MODELS}
\label{section_comparative_axiomatic_analysis}
In order to analytically check how differently the VN method satisfies retrieval constraints as compared to the corresponding original model, we present a comparative axiomatic analysis performed under the retrieval constraints introduced by \citeN{fang04} \footnote{
Note that our goal in this section is to `capture' retrieval heuristics of VN retrieval models, but `not' to refine or improve the standard retrieval constraints of \cite{fang04}.}.


\subsection{Reference Retrieval Constraints}
As in the approach of \cite{clinchant10}, we divide the six standard constraints into two different sets -- \emph{form constraints} (i.e., TFC1, TFC2, and TDC in \cite{fang04}) and \emph{normalization constraints} (i.e., LNC1, LNC2, and TF-LNC in \cite{fang04}). The form constraints specify the desirable restrictions on the ``curve'' of a scoring function. Formally, suppose that $q$ consists of a single word $w$ and $f(\phi(d),q)$ is formulated by $g(x,y)$, where $x$ is $c(w,d)$ and $y$ is $idf(w)$. Then, TFC1, TFC2, and TDC \cite{clinchant10,fang11} correspond to, respectively:
\begin{displaymath}
\frac{\partial g(x,y) }{\partial x} > 0, \mbox{                      }\frac{\partial^2 g(x,y) }{\partial^2 x} < 0, \mbox{                      }\frac{\partial g(x,y) }{\partial y} > 0
\end{displaymath}
It can be easily shown that TFCs and TDC are satisfied for all three normalized functions. This is a natural result, because our normalization only linearly transforms the term frequency and retains the original model, without any change to the basic concepts of the original model.

The normalization constraints describe the necessary properties of a retrieval model for the case in which document-specific quantities such as length, verbosity, and scope are different across documents. According to \cite{fang11}, each normalization constraint can be equivalently described by how the score of a document changes after applying a perturbation operator to the document. We introduce three perturbation operators called \textbf{PAN}, \textbf{PLS}, and \textbf{PAR} that correspond to LNC1, LNC2, and TF-LNC, respectively, as follows\footnote { PAN, PLS, and PAR correspond to TN, LV3, and TG1, respectively, in \cite{fang11}}:

1)	\textbf{PAN} (\textbf{P}erturbation of \textbf{A}dding \textbf{N}oise Words): PAN is an operator for adding noise terms, denoted by $\psi_{AN}$. Given $d$, $\psi_{AN}(d)$ is obtained by adding $K$ noise words $v_1 \cdots v_K$ to $d$, i.e., $\psi_{AN}(d)$ = $d \mbox{ } v_1 \cdots v_K$, where $v_i \notin q$. When $d_2$ = $\psi_{AN}(d_1)$, $| d_2|=|d_1|+K$ and $c(w,d_2)=c(w,d_1)$ for all $w \notin q$.

2)	\textbf{PLS} (\textbf{P}erturbation of \textbf{L}ength \textbf{S}caling): PLS is a length scaling operator, denoted by $\psi_{LS}$. Given $d$, $\psi_{LS}(d)$ is obtained by concatenating \textbf{all query words} in $d$ $K$ times and by scaling the length of $d$ up to $K$ times.
When $d_1$ = $\psi_{LS}(d_2)$, $| d_1| =K \cdot | d_2|$ and $c(w, d_1)=K \cdot c(w,d_2)$ for all $w \in q$. Note that the concatenation is only applied to query words, not necessarily to non-query words. The non-query words in $d$ might (or might not) be kept in $\psi_{LS}(d)$. In the extreme case, all the non-query words do not appear in $\psi_{LS}(d)$, being replaced with other non-query words.

3)	\textbf{PAR} (\textbf{P}erturbation of \textbf{A}dding \textbf{R}elevant Words): PAR is an operator for adding a single relevant word, denoted by $\psi_{AR}$. Given $d$, $\psi_{AR}(d)$ is obtained by appending a single query word $w \in q$, i.e., $\psi_{AR}(d)$ = $d\mbox{ } w \cdots w$ (i.e., the attached number of $w$ is $K$) $K$ times. When $d_1$ = $\psi_{AR}(d_2)$, $| d_1| = |d_2|+K$, $c(w, d_1)= c(w,d_2) + K$ for a given single word $w \in q$, and $c(w', d_1)= c(w',d_2)$ for all $w' \neq w$.\\

Here, $K$ is a perturbation parameter. LNC1, LNC2, and TF-LNC can now be equivalently described as follows:

\textbf{LNC1}: If $d_2$ = $\psi_{AN}(d_1)$, $f(d_1,q)\geq f(d_2,q)$ for $K \geq 1$.

\textbf{LNC2}: If $d_1$ = $\psi_{LS}(d_2)$, $f(d_1,q) \geq f(d_2,q)$ for $K > 1$.

\textbf{TF-LNC}: Let $q$ = $\left\{w\right\}$ be a query with only one term $w$. If $d_1$ = $\psi_{AR}(d_2)$, $f(d_1,q) > f(d_2,q)$ for $K \geq 1$. \\


The perturbation operator PLS for LNC2 is slightly different from the original version of LNC2 \cite{fang04,fang11}.
In the original version, $d_1$ is fully copied to $d_2$, making them identical. In our PLS, only query words are concatenated $K$ times to $d_1$, and no further assumption is made about non-query words. 
Therefore, PLS is the generalized version of the original operator, including the original version as a special case. This generalization does not cause any inconsistency in the known analysis results of LNC2; the analysis results reported in \cite{fang04,fang11} for LNC2 are also still consistently accepted with our PLS operator.
To see the difference more clearly, Algorithm \ref{alg_PLS_operator} summarizes the detailed description of our PLS operator.

\begin{algorithm}
\caption{The detailed procedure of PLS} \label{alg_PLS_operator}
\begin{algorithmic}[1]
\State Step 1) Given $d$, we apply the original PLS operator of \cite{fang04}' to $d$ to obtain $d'$; $d'$ is obtained by simply concatenating all words in $d$ $K$ times
\State Step 2) Given $d'$, $\psi_{LS}(d)$ is obtained after applying the following procedure:
\State Let $d'$ be $w_1 \cdots w_{|d'|}$
\State Initialize $\psi_{LS}(d)$ as an empty document.
\For{$i\gets 1, |d|'$}
\If{$w_i \in q$}
\State $\psi_{LS}(d) \leftarrow \psi_{LS}(d) \mbox{ } w_i$ 
\Else
\State $\psi_{LS}(d) \leftarrow \psi_{LS}(d) \mbox{ } w'$ ($w' \notin q$) where $w'$ is randomly chosen from ${\mathcal{V}} \setminus q$.
\EndIf
\EndFor
\end{algorithmic}
\end{algorithm}


\subsection{Analysis Results of Normalization Constraints}
\label{section_analysis_result_normalization_constraints}

\subsubsection{Assumption}
Before presenting our analysis results of the three normalization constraints, we make the following assumption:
\begin{itemize}
  \item $A_1$: For any query word $w \in q$, $w$ is assumed to be a content-bearing word (i.e., $df(w) \leq N/2$, and $c(w,d) \geq |d|p(w|C)$ for any document $d$ in the collection).
\end{itemize}
Empirically, $A_1$ holds well in usual cases when we filter out stopwords. Table \ref{tbl_C1_test} lists the percentage of $A_1$ being satisfied using all non-stopwords in all queries from three different collections and three query types. (Refer to Section \ref{section_data_set} for a description of the collections and query types.) As shown in Table \ref{tbl_C1_test}, $c(w,d) \geq |d|p(w|C)$ is satisfied in more than 98\% of the documents for all query words in ROBUST, more than 95\% in WT10G, and more than about 93\% in GOV2. The condition $df(w) \leq N/2$ is satisfied for more than 98\% of the query terms.

\begin{table}
\tbl{Percentages of $A_1$ being satisfied using all non-stopwords in all queries from three collections (ROBUST, WT10G, and GOV2) and three query types (sk, sv, and lv).
The columns $A_1^{Okapi}$ and $A_1^{DP}$ indicate the conditions $df(w) \leq N/2$ and $c(w,d) \geq |d|p(w|C)$, respectively.\label{tbl_C1_test}}{%
\begin{tabular}{||c||c|c||c|c||c|c||}\hline
 & \multicolumn{2}{c||}{ROBUST} & \multicolumn{2}{c||}{WT10G} & \multicolumn{2}{c||}{GOV2} \\\hline
  & $A_1^{Okapi}$ & $A_1^{DP}$ & $A_1^{Okapi}$ & $A_1^{DP}$ & $A_1^{Okapi}$ & $A_1^{DP}$ \\\hline
sk & 99.5\% & 99.1\% & 98.5\% & 96.2\% & 99.3\% & 95.6\% \\\hline
sv & 99.4\% & 98.3\% & 99.1\% & 95.1\% & 98.9\% & 93.5\% \\\hline
lv & 99.3\% & 98.3\% & 99.3\% & 95.1\% & 99.2\% & 92.9\% \\\hline
\end{tabular}
}
\end{table}

\subsubsection{Analysis Results}
There exist necessary conditions common for all VN retrieval models under $A_1$ to be satisfied for each normalization constraint. Table \ref{tbl_analysis_result} summarizes the analysis results of the general and the special cases of scope using LengthPower and UniqLength for VN retrieval models, relative to the original models \footnote{Note that the analysis results are obtained from DP and Okapi, not from MRF. For MRF, we do not separately carry out axiomatic analysis, since it is not a base model like DP and Okapi, but being an extension of a base model (i.e. the scoring function of MRF is defined in terms of the main function of its base model). Thus, it is reasonable to assume that the normalization heuristics of MRF will not be significantly different from its base model, without separate analysis.}.


\begin{table}
\tbl{Analysis results of the original and VN retrieval models for three normalization constraints -- LNC1, LNC2, and TF-LNC -- under $A_1$.\label{tbl_analysis_result}}{%
\begin{tabular}{||c|c|c|c||}\hline
 & LNC1 & LNC2 & TF-LNC \\\hline
Original \cite{fang04} & Yes & Yes & Yes \\\hline
Verbosity-normalized & $C_1$ & $C_2$  & $C_3$ \\
(General) & & & \\\hline
Verbosity-normalized & Yes & Yes & Yes \\
 (LengthPower) &  &  &  \\\hline
Verbosity-normalized & $C_1$ & $C_2$ & Yes \\
 (UniqLength) &  &  &  \\\hline
Verbosity-normalized & $C_1$ & $C_2$  & $C_4$ \\
 (EntropyPower) & & & \\\hline
\end{tabular}
}
\end{table}

Table \ref{tbl_analysis_result} uses the notations introduced by \cite{fang04}, where ``Yes'' and ``$C_x$'' indicate that the corresponding model satisfies the particular constraint in the absence of conditions and under particular conditions, respectively. The specific conditions are


$\bullet$ $C_1$: $v(d_2) \geq v(d_1)$ \\

$\bullet$ $C_2$: $s(d_1) \geq s(d_2)$ \\

$\bullet$ $C_3$: $K/c(w,d_2) \geq v(d_1)/v(d_2)-1$ \\

$\bullet$ $C_4$: $s(d_2) \leq \left( |d_2| /c(w,d_2)\right)^2$ \\

where  $C_1$, $C_3$ and $C_4$ are sufficient but not necessary conditions to satisfy the particular constraint. Some derivations of the conditions are given in Appendix C-E.

As shown in Table \ref{tbl_analysis_result}, an original method
satisfies all three constraints unconditionally under $A_1$ according
to \cite{fang04}, whereas a VN method requires additional conditions
that depend on the choice of scope measure. An exceptional case is
LengthPower, in which all constraints are satisfied unconditionally.

\begin{table}
\tbl{Percentages of $C_4$ being satisfied using all non-stopwords in all queries from three collections (ROBUST, WT10G, and GOV2) and three query types (sk, sv, and lv).\label{tbl_C4_test}}{%
\begin{tabular}{||c|c|c|c||}\hline
 & ROBUST & WT10G & GOV2 \\\hline
sk & 99.99\% & 99.97\% & 99.98\% \\\hline
sv & 99.99\% & 99.98\% & 99.98\% \\\hline
lv & 99.99\% & 99.98\% & 99.96\% \\\hline
\end{tabular}
}
\end{table}

Among the three constraints, TF-LNC is satisfied under LengthPower and UniqLength, the detailed proofs of which are presented in Appendix D.
Under EntropyPower, TF-LNC is satisfied for almost all query words in our test collection, as shown in Table \ref{tbl_C4_test}; $C_4$ is satisfied in more than 99.9\% of the documents for all query words in ROBUST, WT10G and GOV2. Therefore, we do not explore TF-LNC further in this paper.




\subsection{Normalization Heuristics of VN Retrieval Models (Case: UniqLength and EntroyPower)}
\label{section_analysis_heuristic}

In this section, we discuss the retrieval behaviors entailed
from the VN method in the cases of UniqLength and EntropyPower, with respect to the original method.
In our discussion, PAN and PLS are further divided into two different types -- V-type and S-type -- which refer to  \emph{verbosity-increasing} and \emph{scope-broadening} perturbations, respectively. The definitions of these types of operators are as follows:
\begin{enumerate}
  \item  \textbf{V-type perturbation}: The operator $\psi(\cdot)$ is called \emph{V-type} if the perturbation does not \emph{increase} the scope of the document, i.e., if $d_1$ = $\psi(d_2)$ and $\psi$ is V-type, $s(d_1) \leq s(d_2)$.
  \item  \textbf{S-type perturbation}: The operator $\psi(\cdot)$ is called \emph{S-type} if the perturbation does not \emph{decrease} the scope of the document, i.e., if $d_1$ = $\psi(d_2)$ and $\psi$ is S-type, $s(d_1) \geq s(d_2)$
\end{enumerate}
We then reexamine how the original and VN models satisfy LNCs on V-type and S-type PAN and PLS.

The notable result is that $C_1$ and $C_2$ correspond to a relaxed penalization of a scope-broadened document, and a strict penalization of a verbosity-increased document, respectively.

First, we present the first heuristic H1 and discuss its derivation from $C_1$:

\subsubsection{\textbf{H1: Relaxed penalization of scope-broadened documents}}
\emph{The VN retrieval method performs a relaxed penalization of a scope-broadened document after performing PAN. (from LNC1) }

To derive H1, we divide PAN into V-PAN and S-PAN. \textbf{V-PAN} denotes \emph{verbosity-increasing} PAN, where $K$ added noise words are covered by the original scope of the document. \textbf{S-PAN} denotes the \emph{scope-broadening} PAN, where $K$ added noise words describe \emph{new} contents that are not covered by the scope of the original document. In terms of V-type and S-type, V-PAN and S-PAN can be defined as follows:
\begin{enumerate}
  \item  \textbf{V-PAN}: V-PAN is a specific type of PAN, being V-type.
  \item 	\textbf{S-PAN}: S-PAN is a specific type of PAN, being S-type.
\end{enumerate}

Suppose that $d_1$ and $d_2 = \psi_{AN}(d_1)$ are the given documents for LNC1. Then, we can show that the VN model often does \emph{not} penalize $d_2$ for S-PAN; instead, it penalizes $d_2$ for V-PAN. On the other hand,  the original model \emph{always} penalizes $d_2$ for both S-PAN and V-PAN. Thus, the VN model imposes a type of relaxed penalization to \emph{scope-broadened} documents after PAN, with respect to the original model.

Equivalently, the heuristic H1 can be rewritten in the form of a retrieval constraint as follows:

\textbf{H1-LNC}: If $d_2$ = $\psi_{AN}(d_1)$ and $\psi_{AN}$ is S-PAN, $f(d_1, q) \leq f(d_2, q)$ for $K \geq 1$ with the following condition $C_5$ and $C_6$, for VN-Okapi and VN-DP, respectively:

$\bullet$ $C_5$:
  \begin{equation}
  \frac{v(d_1) - v(d_2)}{K } \geq \frac{b}{1-b} \frac{1}{avgs}
\label{eq_C5_original}
  \end{equation}

$\bullet$ $C_6$:
\begin{equation}
  \frac{v(d_1) - v(d_2)}{K } \geq \frac{1}{s(d_1)} \frac{p(w|C) + p_{ml}(w|d) s(d_1) \mu^{-1}}{p_{ml}(w|d) - p(w|C)}
  \label{eq_C6_original}
\end{equation}
where $p(w|d) > p(w|C)$ is additionally assumed in $C_6$\footnote{For $C_6$, assuming an extreme case where the query is a very highly topical, i.e., $p_{ml}(w|d) \gg   p(w|C)$ or $r \rightarrow \infty$ ), $C_6$ is simplified as:
\begin{equation}
  \frac{v(d_1) - v(d_2)}{K } \geq \frac{1}{\mu}
\label{eq_C6_simplified}
\end{equation}
}.


Compared to LNC1, the consequence part of H1-LNC conditionally entails the negation of LNC1, implying that the VN model often \emph{prefers} some of the scope-broadened documents resulting from S-PAN, although the original model does not \footnote{From Eq. (\ref{eq_C5_original}) and Eq. (\ref{eq_C6_original}), when $\left(v(d_1)-v(d_2)\right)/K$ is sufficiently large, $C_5$ (or $C_6$) can be satisfied both for VN-Okapi and VN-DP. This case can appear if $v(d_1)$ is large and $s(d_1)$ is small, but it is not always true. Otherwise, $C_5$ (or $C_6$) can be satisfied, according to the choice of a retrieval parameter value or a term discrimination value of a query word; for VN-Okapi, $C_5$ is satisfied if $b$ is sufficiently large; for VN-DP, $C_6$ is satisfied if $\mu$ is sufficiently large and the query word is highly topical. 
}.

$\bullet$ \emph{Example of H1}: Here, we present examples of S-PAN and V-PAN. Suppose that we use UniqLength as the scope measure and a document consisting of passages that are disjoint in scope. Formally, let ${\mathbi{g}}$, $h$, and $x$ be passages, and assume that ${\mathbi{g}}$, $h$, and $x$ have no common or overlapping content, where ${\mathbi{g}}$ denotes a relevant passage and $h$ and $x$ are non-relevant, i.e., $c(w, {\mathbi{g}})$ $>$ 0, $c(w, h)$ = $c(w, x)$ = 0 for query word $w \in q$. Examples of S-PAN and V-PAN are as follows:

\begin{minipage}[t]{1.0\linewidth}
\textbf{Example of S-PAN:}\\
\framebox{
\parbox{230pt}{
$d_1$ = ${\mathbi{g}}$ ${\mathbi{g}}$ $h$ $h$ \\
$d_2$ = ${\mathbi{g}}$ ${\mathbi{g}}$ $h$ $h$ $x$
}}

\textbf{Example of V-PAN:}\\
\framebox{
\parbox{230pt}{
$d_1$ = ${\mathbi{g}}$ ${\mathbi{g}}$ $h$ $h$ \\
$d_2$ = ${\mathbi{g}}$ ${\mathbi{g}}$ $h$ $h$ $h$
}}
\end{minipage}

For both examples, the query relevant content is not changed after PAN. Because these two examples are PAN examples, the original method always prefers $d_1$ to $d_2$, irrespective of the PAN type. However, $C_1$ is satisfied only for V-PAN, because $s(d_2)$ = $s(d_1)$ and $v(d_2) \geq v(d_1)$, and not clearly for S-PAN, because $v(d_2) < v(d_1)$ is plausible due to $s(d_2) > s(d_1)$. Therefore, the VN method prefers $d_2$ in V-PAN and not always in S-PAN.

$\bullet$ \emph{Derivation of H1}: To show how the VN model behaves differently toward V-PAN and S-PAN, we first rewrite $C_1$ by $s(d_2) - s(d_1)$ $\leq$ $K/v(d_1)$, implying that the scope of the new document $d_2$ must not be increased considerably after performing PAN.

\emph{i) Case: V-PAN}

First, V-PAN does not increase $s(d_2)$ according to the definition of a V-type perturbation, resulting in $s(d_2) -s(d_1) \leq 0$. As a result, it is clear that $C_1$ is always true for V-PAN, finally making LNC1 true. Thus, for V-PAN, there is no difference between the original and the VN models in satisfying LNC1.

For example, suppose that we use UniqLength as the scope measure and consider a given V-PAN in which all $K$ words already occur in $d_1$. In this case, $s(d_2) = s(d_1)$ because no new words occur in $d_2$. Thus, $C_1$ is equivalent to $s(d_2) - s(d_1)$= $0 \leq K/v(d_1)$, which is true irrespective of $K$.

\emph{ii) Case: S-PAN}

Second, S-PAN increases the scope after performing PAN according to the definition of an S-type perturbation, resulting in $s(d_2) - s(d_1) \geq 0$. Therefore, $C_1$ is not always true.

For example, suppose that we use UniqLength as the scope measure again, and consider a given S-PAN in which all $K$ words are new and different from each other. In this case, $s(d_2)$ = $s(d_1) + K$, and $C_1$ is equivalent to $K \leq K/v(d_1)$; however, $C_1$ is usually not satisfied because $v(d_1) \geq 1$.

Instead, it often satisfies the \emph{negation} of LNC1. 
Consider the same S-PAN example in which all $K$ words are different from each other and assume that we use VN-DP as an example retrieval model. 
$C_6$ is then equivalent to:
\begin{equation}
  \frac{v(d_1) - 1}{1+ K/s(d_1) } \geq  \frac{p(w|C) + p_{ml}(w|d) s(d_1) \mu^{-1}}{p_{ml}(w|d) - p(w|C)}
\label{eq_C6_simplified_example}
\end{equation}

There exist a number of situations in which $C_6$ is true according to Eq. (\ref{eq_C6_simplified_example}) (i.e., $v(d_1)$ is sufficiently large, or $K$ added words are highly topical ($p(w|d) \gg p(w|C)$) and $\mu$ is reasonably large). Therefore, for S-PAN, the VN model often does not satisfy LNC1, in contrast to the original model that always satisfies LNC1 $\square$.

Next, we present the heuristic H2 and discuss its derivation from $C_2$:

\subsubsection{\textbf{H2: Strict penalization of verbosity-increased documents}}
\emph{The VN retrieval method imposes a strict penalization of a verbosity-increased document after performing PLS (from LNC2).}

As was performed on PAN, we divide PLS into V-PLS and S-PLS. \textbf{V-PLS} denotes \emph{verbosity-increasing} PLS, where the non-query words after PLS is performed are covered by the original scope of the document, thereby increasing verbosity. \textbf{S-PLS} denotes the \emph{scope-broadening} PLS, where the non-query words after PLS is performed introduce \emph{new} contents that are not covered by the scope of the original document. In terms of V-type and S-type, V-PLS and S-PLS can be defined as follows:
\begin{enumerate}
  \item  \textbf{V-PLS}: V-PLS is a specific type of PLS, being V-type.
  \item  \textbf{S-PLS}: S-PLS is a specific type of PLS, being S-type.
\end{enumerate}

Given two documents $d_1  =\psi_{LS}(d_2) $ and $d_2$ for LNC2, from the definition of $C_2$ (i.e., $s(d_1) \geq s(d_2)$), LNC2 is satisfied only if the scope of the original document increases after PLS is performed. Therefore, the VN model \emph{prefers} (or does not penalize) only $d_1$ for S-PLS because it increases the scope; instead, it \emph{penalizes} $d_1$ for V-PLS, which decreases the scope. As such, the VN model imposes a strict penalization of a verbosity-increased document after PLS.

Equivalently, the heuristic H2 can be rewritten in a form of retrieval constraint as follows:

\textbf{H2-LNC}: If $d_1$ = $\psi_{LS}(d_2)$ and $\psi_{LS}$ is V-PLS, $f(d_1,q) \leq f(d_2,q)$ for $K > 1$.

Compared to LNC2, the consequence part of H2-LNC is the negation of that of LNC2, implying that the VN model \emph{always} penalizes verbosity-increased documents resulting from S-PLS, although the original model does \emph{not} (i.e, prefers them).

$\bullet$ \emph{Example of H2}:  We present examples of S-PLS and V-PLS. Suppose that we use UniqLength as the scope measure and a document consisting of passages that are disjoint in scope. Formally, let ${\mathbi{g}}$, $h$, $x$, and $y_i$ be passages with equal length (i.e., $|{\mathbi{g}}|$ = $|h|$ = $|x|$ = $|y_i|$) and unit scope (i.e., $s(g)$ = $s(h)$ = $s(x)$ = $s(y_i)$ = 1), and assume that ${\mathbi{g}}$, $h$, $x$, and $y_i$ have no common or overlapping content, where ${\mathbi{g}}$ denotes a relevant passage and $h$, $x$, and $y_i$ are non-relevant, i.e., $c(w,{\mathbi{g}}) > 0$, $c(w,h)$ = $c(w,x)$ = $c(w,y_i)$ = 0 for query word $w \in q$. Examples of S-PLS and V-PLS are as follows:

\textbf{Example of S-PLS:}\\
\framebox{
\parbox{230pt}{
$d_1$ = ${\mathbi{g}}$ ${\mathbi{g}}$ $h$ $x$ $y_1$ $y_2$ \\
$d_2$ = ${\mathbi{g}}$ $h$ $x$
}}

\textbf{Example of V-PLS:}\\
\framebox{
\parbox{230pt}{
$d_1$ = ${\mathbi{g}}$ ${\mathbi{g}}$ $h$ $h$ $h$ $h$\\
$d_2$ = ${\mathbi{g}}$ $h$ $x$
}}

For both examples, $|d_2|$ = $2|d_1|$, $c(w,d_1)$ = $2 c(w,d_2)$ for $w \in q$, i.e., the query-relevant content is copied twice. The example of S-PLS introduces two \emph{new} non-relevant passages $y_1$ and $y_2$ that are not given in $d_2$, whereas the example of V-PLS does not introduce any new passage but only repeats the previously mentioned non-relevant passage $h$.

Because these two examples are PLS examples, the original method always prefers $d_1$ to $d_2$, irrespective of the PLS type. However, $C_2$ is satisfied only for S-PLS and not for V-PLS, from $s(d_1) = 5 > s(d_2) = 3$ in S-PLS and $s(d_1) = 2 < s(d_2) = 3$ in V-PLS. Therefore, the VN method prefers $d_1$ only in S-PLS and not in  V-PLS.

$\bullet$ \emph{Derivation of H2}:  From the definitions of V-type and S-type perturbations, it is trivial to show that $C_2$ is true for V-PLS but false for S-PLS. Therefore, for S-PLS, the VN model does not satisfy LNC2, in contrast to the original model which always satisfies LNC2 $\square$.

\subsubsection{Summary}

Table \ref{tbl_analysis_result_of_four_perturbations} summaries the normalization behaviors of the original and VN models in response to S-PAN, V-PAN, S-PLS, and V-PLS.

For V-PAN and S-PLS, the VN model leads to the same normalization heuristics as those of the original model. For S-PAN and V-PLS, however, the normalization behaviors are completely different between the original and the VN models; for S-PAN, the VN model often does \emph{not} penalize the new document, whereas the original model \emph{always} penalizes it; for V-PLS, the VN model \emph{always} penalizes the new document, whereas the original model does \emph{not} penalize it.

Overall, the normalization heuristics entailed from the VN model are dependent on whether a perturbation is V-type or S-type. For a V-type perturbation, the VN model imposes a strict penalization of a verbosity-increased document (i.e., entailing H1), irrespective of PAN or PLS. For an S-type perturbation, the VN model unlikely penalizes a scope-broadened document (i.e., entailing H2). On the other hand, the normalization heuristics of the original model are dependent on whether a perturbation is PAN or PLS, not on whether it is V-type or S-type. For PAN, the original model penalizes a new document, irrespective of whether the document is verbosity-increased or scope-broadened. For PLS, it does not penalize the new document.

\begin{table}
\tbl{The summary of the normalization behaviors of the original and VN models for four perturbations -- S-PAN, V-PAN, S-PLS, and V-PLS -- under $A_1$
in which $d$ is original document, and $\psi_{AN}(d)$ (or $\psi_{LS}(d)$) is the perturbed documents of $d$ after PAN (or PLS). \label{tbl_analysis_result_of_four_perturbations}}{%
\begin{tabular}{||l|c|c||}\hline
$\psi$&  Verbosity-normalized model & Original model \\
    &  (UniqLength, EntropyPower)  & \\\hline
S-PAN & $\begin{array}{lc} f(d,q) \leq f(\psi_{AN}(d),q) & \mbox{if } C_5 \mbox{(or } C_6\mbox{) is true } \\ f(d,q) \geq f(\psi_{AN}(d),q) & \mbox{Otherwise}  \end{array}$ & \multirow{2}{*}{$f(d,q) \geq f(\psi_{AN}(d),q)$} \\\cline{1-2}
V-PAN & $f(d,q) \geq f(\psi_{AN}(d),q)$ &  \\\hline
S-PLS & $f(d,q) \leq f(\psi_{LS}(d),q)$ & \multirow{2}{*}{$f(d,q) \leq f(\psi_{LS}(d),q)$} \\\cline{1-2}
V-PLS & $f(d,q) \geq f(\psi_{LS}(d),q)$ & \\\hline
\end{tabular}
}
\end{table}

\section{EXPERIMENTAL SETTING}
\label{section_experiments}

\subsection{Experimental Setup}
\label{section_data_set}

For evaluation, we used three different standard TREC collections --
ROBUST, WT10G, and GOV2. Table \ref{tbl_collection_statistic} lists
the basic statistics of each test collection, where \emph{NumDocs} is
the number of documents, \emph{NumWords} is the total number of word
occurrences in each collection, \emph{TopicSet} is the range of topic
numbers used for training and testing, and \emph{Avg} of $|d|$,
$h(d)$, and $v(d)$ indicates the average length, entropy power, and
verbosity\footnote{Entropy power is used as scope measure for $v(d)$.}, respectively, in a given collection.  ${\mathsf{CoeffVar}}$
is the corresponding \emph{coefficient of variance}, which is defined
as the ratio of the standard deviation to the mean. The interesting
statistic is ${\mathsf{CoeffVar}}$ of $v(d)$, which indicates the differences among
the verbosities of documents in a collection.  ROBUST has
the most similar verbosities across documents, whereas GOV2 has the
most different verbosities. This is because many documents in ROBUST are
newspaper documents, for example, from Financial Times and Los Angeles
Times, which are more homogeneous collections.  In contrast, the web
documents in GOV2 are more heterogeneous.

\begin{table}
\tbl{Statistics of each test collection. ${\mathsf{CoeffVar}}$ denotes the coefficient of variation.\label{tbl_collection_statistic}}{%
\begin{tabular}{||l|c|c|c||}\hline
Statistics & ROBUST & WT10G & GOV2 \\\hline
\emph{NumDocs} & 528,156 & 1,692,096 & 25,205,179 \\\hline
\emph{NumWords} & 572,180 & 6,346,858 & 40,002,579 \\\hline
\emph{TopicSet} & Q301$-$450 & Q451$-$550 & Q701$-$850 \\
    & Q601$-$700 & & \\\hline
\emph{Avg of} $|d|$ & 233.34 & 400.25 & 690.8 \\
(\emph{${\mathsf{CoeffVar}}$}) & (2.39) & (6.06) & (2.86) \\\hline
\emph{Avg of} $h(d)$ & 107.77 &  109.60  & 109.85 \\
(\emph{${\mathsf{CoeffVar}}$}) & (0.81) & (1.45) & (0.98) \\\hline
\emph{Avg of} $v(d)$ & 1.77 & 2.95 &  6.11 \\
(\emph{${\mathsf{CoeffVar}}$}) & (0.91) & (5.51) & (7.17) \\\hline
\end{tabular}
}
\end{table}

\sloppypar{
All experiments were performed using the Lemur toolkit (version 4.12). 
We carried out standard preprocessing by applying the Porter stemmer and removing stopwords from the standard INQUERY stoplist \cite{inquery00}. 
To cover different types of queries, we follow the setting used in \cite{zhai01}, where four combinations are used: short keywords (\textbf{sk}, title), short verbose (\textbf{sv}, description), long keywords (\textbf{lk}, concept), and long verbose (\textbf{lv}, title, description, and narrative). In our test topic sets, because lk is not available, the other three types were examined. We use MAP (mean average precision) and P@5 (precision at top 5 documents) as the evaluation measures \cite{croft09}.

For each query, our evaluation is based on the top 1,000 documents retrieved. We also report significance test results by a non-directional paired t-test at 0.95 confidence level.
For the significance test, we use all \emph{per-topic} performances in a collection, i.e., the number of performance difference samples used for the t-test is the same as the total number of topics in a given collection
\footnote{ In Section \ref{section_parameter_training}, we introduce a K-fold cross-validation to avoid the optimization of the retrieval parameters to the test set.
However, note that we do not use \emph{per-fold} performances to perform the significance test but simply use all per-topic performances. To the best of our knowledge, this type of significance test is an IR-specific setting that is different from the other types of significance test used in non-IR literatures.

}.

\subsection{Parameter Tuning}
\label{section_parameter_training}

Several tuning parameters are present in the retrieval methods -- DP:
$\mu$ and Okapi: $b$, $k_1$, and $k_3$. Given a test topic set
consisting of 50 queries, each parameter was tuned using the other
topic sets in the same test collection as the development set\footnote{Here, a topic set consists of 50 queries, which was were created in each year of TREC. For example, in ROBUST, as 250 queries are available, there are 5 topic sets, namely, TREC6(Q301$-$Q350), TREC7(Q351$-$Q400), TREC8(Q401$-$Q450), ROBSUT03(Q601$-$Q650), and ROBUST04(Q651$-$Q700). Parameters used when testing 50 queries in each topic set are trained using the other 200 queries in other topic sets as training data.  In other words, for testing 50 queries in TREC6, queries Q351$-$Q450 and Q601$-$Q700 in TREC7, TREC8, ROBUST03, and ROBUST04 are used as training data. For testing queries in TREC7, queries in TREC6, TREC8, ROBUST03, and ROBUST04 are used as the training set, and so on. Therefore, for ROBUST, we use a five-fold cross validation for parameter tuning, whose folds are fixed.
For WT10G, where 100 queries are used, we have two topic sets, namely TREC9(Q451$-$Q500) and TREC10(Q501$-$Q550). For testing 50 queries in TREC9, we use queries in TREC10 as the training set, and vice versa. Thus, for WT10G, we use a two-fold cross validation for parameter tuning. 
Similarly, for GOV2, 150 queries are available, so we have three topic sets, namely TREC2004(Q701$-$Q750), TREC2005(Q751$-$Q800) and TREC2006(Q801$-$Q850). For testing 50 queries in TREC2004, queries in TREC2005 and TREC2005 are used as the training data.  Thus, for GOV2, we use a three-fold cross validation for parameter tuning. 
}. The search space for each parameter is given as follows:\\
$\bullet$ $\mu$: \{ 100, 200, 300, 400, 500, 600, 800, 1000, 1500, 2000, 2500, 3000, 4000, 5000, 7000, 10000, 15000, 20000 \}\\
$\bullet$ $b$: \{0, 0.001, 0.003, 0.005, 0.007, 0.01, 0.02, 0.03, 0.05, 0.1, 0.2, 0.3, 0.4, 0.5, 0.6, 0.7, 0.8, 0.9\}\\
$\bullet$ $k_1$: \{0.25, 0.3, 0.4, 0.5, 0.6, 0.8, 1.0, 1.2, 1.5, 1.8, 2.0, 2.5, 3.0\}\\
$\bullet$ $k_3$: fixed at 1,000\\


In our preliminary experiments, we found that LengthPower for $s(d)$ can suffer from the parameter scaling problem, in which the optimal parameter ranges of $\mu$ and $k_1$ in the VN methods differ from the known ranges of the original model. For instance, when $\beta = 0.25$, it was found that $\mu$ was optimal at a value of less than 100, which is beyond the normal parameter range.
To resolve the scaling problem, we substitute $avgv$ for $k$, instead of setting $k$ to 1, such that $c(w,\phi(d))$ would become $c(w,d)$ on average. This consideration leads to the following parameter scaling:
\begin{displaymath}
k_1 \leftarrow k_1 \cdot avgv^{-1},\mbox{           } \mu \leftarrow \mu \cdot avgv^{-1}
\end{displaymath}
This parameter scaling is applied only to LengthPower, and not to the others. No such parameter scaling problem occurs in the case of UniqLength and EntropyPower.



\section{Experimental Results}
\label{section_experimental_results}

 This section reports the comparative results of the original and the VN retrieval method for Okapi, DP and MRF.

\subsection{DP vs. VN-DP}
\label{section_expr_dp}

Table \ref{tbl_experiment_result_VN_DP} show the comparative results (MAP) of DP and VN-DP under three different scope measures, UniqLength, EntropyPower, and LengthPower($\beta$), which are denoted as $l_\beta(d)$, $u(d)$, and $h(d)$, respectively.

\begin{table}
\tbl{MAP performance comparison of DP and VN-DP on three collections ROBUST, WT10G, and GOV2;
three different query types sk, sv, and lv; and three different scope measures  LengthPower ($\beta$), UniqLength, and EntropyPower.
The row titled ``baseline'' indicates the original model. The symbols * indicate that a run of the VN method shows statistically significant improvement over the baseline in the t-test, at 0.95 confidence level.\label{tbl_experiment_result_VN_DP}}{%
\begin{tabular}{||l|c||c|c|c||}\hline
 & Method & \multicolumn{3}{c||}{DP (or VN-DP)} \\\cline{3-5}
 &  & ROBUST & WT10G & GOV2 \\\hline
\multirow{6}{*}{sk} & baseline & 0.2447 & 0.1963 & 0.2907 \\\cline{2-5}
 & LengthPower(0.25) & 0.2252 & 0.1649 & 0.2403 \\\cline{2-5}
 & LengthPower(0.5) & 0.2401 & 0.1953 & 0.2823 \\\cline{2-5}
 & LengthPower(0.75) & 0.2457 & 0.1968 & 0.2930 \\\cline{2-5}
 & LengthPower(0.9) & 0.2460* & 0.1963 & 0.2913 \\\cline{2-5}
 & UniqLength & 0.2472* & 0.2046  & 0.3055* \\\cline{2-5}
 & EntropyPower & \textbf{0.2481}* & \textbf{0.2120}* & \textbf{0.3099}* \\\hline\hline
\multirow{6}{*}{sv} & baseline & 0.2260 & 0.1909 & 0.2455 \\\cline{2-5}
 & LengthPower(0.25) & 0.2312 & 0.1790 & 0.2350 \\\cline{2-5}
 & LengthPower(0.5) & \textbf{0.2443}* & 0.2103* & 0.2633* \\\cline{2-5}
 & LengthPower(0.75) & 0.2396* & 0.2044* & 0.2569* \\\cline{2-5}
 & LengthPower(0.9) & 0.2319* & 0.1946* & 0.2487* \\\cline{2-5}
 & UniqLength & 0.2385* & 0.2109* & 0.2671* \\\cline{2-5}
 & EntropyPower & 0.2440* & \textbf{0.2196}* & \textbf{0.2826}*  \\\hline\hline
\multirow{6}{*}{lv} & baseline & 0.2707 & 0.2469 & 0.2864 \\\cline{2-5}
 & LengthPower(0.25) & 0.2697 & 0.2249 & 0.3060* \\\cline{2-5}
 & LengthPower(0.5) & 0.2765* & 0.2506 & 0.3133* \\\cline{2-5}
 & LengthPower(0.75) & 0.2762* & 0.2532* & 0.3005* \\\cline{2-5}
 & LengthPower(0.9) & 0.2725* & 0.2501* & 0.2914* \\\cline{2-5}
 & UniqLength & 0.2759* & 0.2553* & 0.3083* \\\cline{2-5}
 & EntropyPower & \textbf{0.2799}* & \textbf{0.2614}* & \textbf{0.3248}* \\\hline
\end{tabular}
}
\end{table}

Generally, it is observed that VN-DP improves original DP. These improvements are
statistically significant for almost all test collections and all
query types (for both UniqLength and EntropyPower), often resulting in
an improvement of 10\%. The improvement tends to be larger on Web collections (i.e., WT10G and GOV2) than for ROBUST.  A possible reason is that the Web collections
have higher ${\mathsf{CoeffVar}}$ of $v(d)$ because of the heterogeneity
of documents, and thus, they could gain more from our verbosity
normalization.


Among the three scope measures, EntropyPower is the best, and it outperforms UniqLength and
LengthPower for most topic sets. UniqLength is slightly better than
LengthPower; however, the difference in their performances is not
significant. When the best $\beta$ value is adopted for each query
type, LengthPower can often show performance similar to that of UniqLength.


Interestingly, VN-DP leads to significant improvements, more in
verbose queries (i.e., sv and lv) than in keyword queries.  For
EntropyPower, VP-DP causes an improvement of 1.55\% in ROBUST for
short keyword queries, 8.2\% in WT10G, and 6.64\% in GOV2. The
corresponding improvements are much larger for short verbose queries,
being 8.05\% in ROBUST, 16.66\% in WT10G, and 15.11\% in GOV2, and they
are also large for long verbose queries. Restricting our discussion to
DP, these results strongly support that the use of heuristics H1 and H2
is indeed important, especially for
verbose queries.

In addition, Table \ref{tbl_experiment_result_P5_VN_DP} shows the performances of P@5 for VN-DP, as compared to that of DP, based on the MAP-optimized free-parameters' values used in Table \ref{tbl_experiment_result_VN_DP}. One reason for using the same retrieval parameters instead of directly optimizing P@5 is that the concavity of the performance curve was smoother in MAP than in P@5, thereby avoiding the use of far-from-optimal parameter values. As similarly mentioned by \cite{kurland09}, this choice helps us to examine whether the improved performance in MAP causes severe degradation or significant improvement in the precision, which is an often important metric in some IR applications such as Web search.

\begin{table}
\tbl{P@5 performance comparison of DP and VN-DP for three collections -- ROBUST, WT10G, and GOV2. \label{tbl_experiment_result_P5_VN_DP}}{%
\begin{tabular}{||l|c||c|c|c||}\hline
 & Method & \multicolumn{3}{c||}{DP (or VN-DP)} \\\cline{3-5}
 &  & ROBUST & WT10G & GOV2 \\\hline
\multirow{6}{*}{sk} & baseline & 0.4924 & 0.3120 & 0.5678 \\\cline{2-5}
 & LengthPower(0.5) & 0.4707 & 0.3360 & 0.5409 \\\cline{2-5}
 & LengthPower(0.75) & 0.4851 & 0.3220 & 0.5611 \\\cline{2-5}
 & LengthPower(0.9) & 0.4924 & 0.3080 & 0.5691 \\\cline{2-5}
 & UniqLength & 0.4956 & 0.3620* & 0.5906 \\\cline{2-5}
 & EntropyPower & \textbf{0.4972} & \textbf{0.3640}* & \textbf{0.6416}*  \\\hline\hline
\multirow{6}{*}{sv} & baseline & 0.4466 & 0.3880 & 0.5208 \\\cline{2-5}
 & LengthPower(0.5) & 0.4811* & 0.4000 & 0.5383 \\\cline{2-5}
 & LengthPower(0.75) & 0.4699* & 0.3960 & 0.5409 \\\cline{2-5}
 & LengthPower(0.9) & 0.4530 & 0.3820 & 0.5275 \\\cline{2-5}
 & UniqLength & 0.4755* & 0.4060 & 0.5826* \\\cline{2-5}
 & EntropyPower & \textbf{0.4932}* & \textbf{0.4300}* & \textbf{0.6309}*  \\\hline\hline
\multirow{6}{*}{lv} & baseline & 0.5414 & 0.4460 & 0.6228 \\\cline{2-5}
 & LengthPower(0.5) & 0.5518 & 0.4660 & 0.6188 \\\cline{2-5}
 & LengthPower(0.75) & 0.5510 & 0.4560 & 0.6295 \\\cline{2-5}
 & LengthPower(0.9) & 0.5526* & 0.4520 & 0.6282 \\\cline{2-5}
 & UniqLength & 0.5542* & 0.4560 & 0.6456* \\\cline{2-5}
 & EntropyPower & \textbf{0.5631}* & \textbf{0.4700} & \textbf{0.6644}*  \\\hline
\end{tabular}
}
\end{table}

Under EntropyPower, the improvement of P@5 from DP to VN-DP is significant in most cases, often being larger than that of MAP for short keyword and verbose queries. VP-DP improves over DP by about 16.67\% on WT10G and about 13.00\% on GOV2 for short keyword queries, and 13.00\% on WT10G and 21.14\% on GOV2 for short verbose queries. Exceptional cases are found for keyword queries on ROBUST and for long verbose queries on WT10G, where the improvement in the precision is not statistically significant. Therefore, at least for EntropyPower, the results imply that the significant improvements of MAP using VN-DP are caused by the increased performance in P@5. For UniqLength, however, the impact of P@5 and its contribution to MAP is less clear than that for EntropyPower, although some noticeable improvements are observed. For LengthPower, most improvements of P@5 are not statistically significant. This implies that performance metrics other than P@5, such as recall, might be the major factors causing the significant improvement in MAP for LengthPower.

For further comparison, Figure \ref{fig_DP_performance_curve} shows
the performance curves of the original DP and VN-DP using
EntropyPower, plotted by varying $\mu$ for MAP and P@5. For both measures,
VN-DP is always better than DP, for almost all $\mu$ values and in all test collections and query
types. The shapes of the curves of DP and VN-DP are similar, and the
optimal ranges of $\mu$ are also fairly similar. This similarity
between DP and VN-DP is also observed in the case of UniqLength,
although we do not present the curves for this scope measure here,
in the interest of conciseness.

\begin{figure}[ht]
\centering
\subfigure[ROBUST (MAP)]{
     \label{fig_DP_performance_curve_a}
     \includegraphics[width=0.45\linewidth]{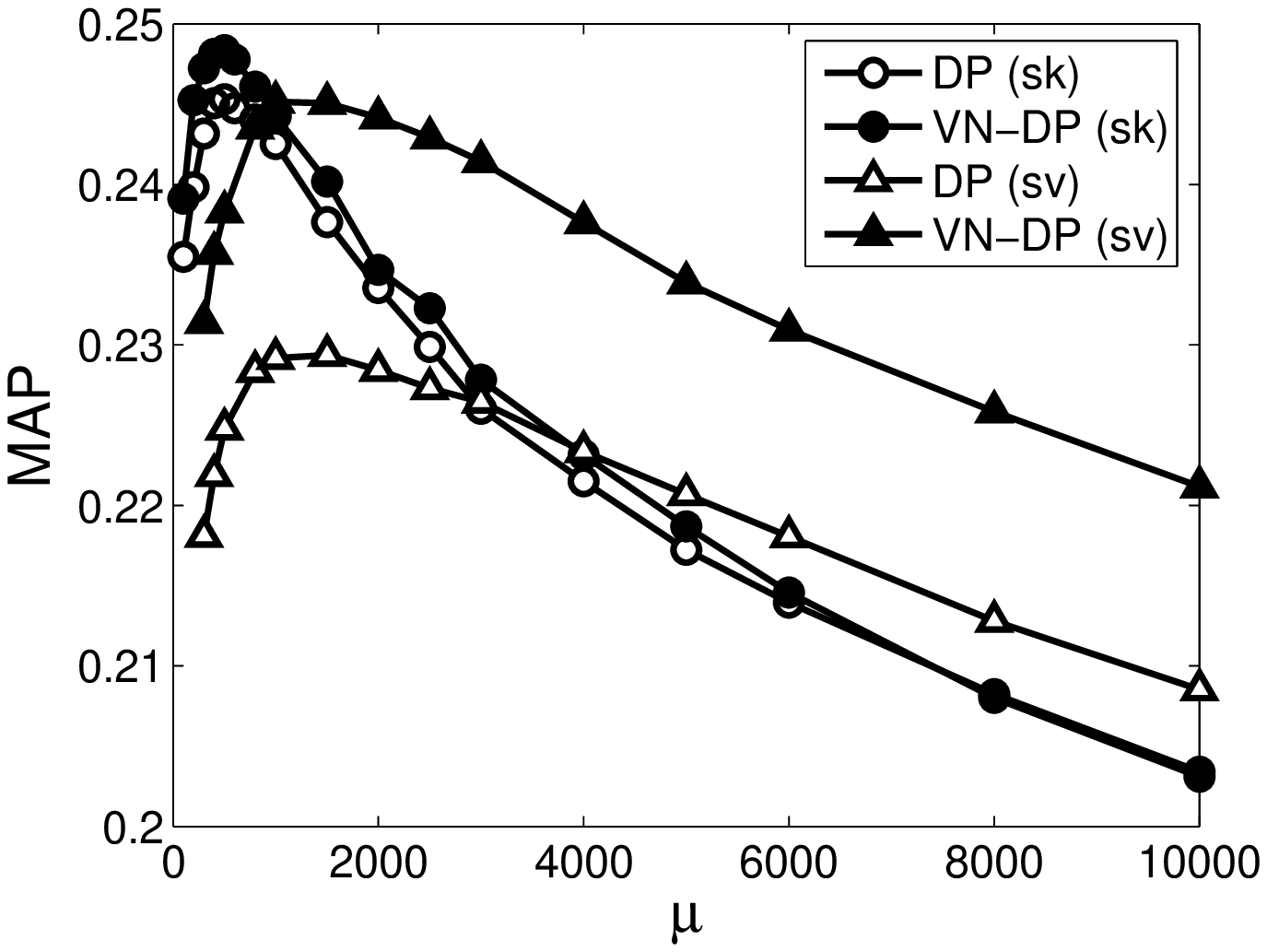}
}
\subfigure[ROBUST (P@5)]{
     \label{fig_DP_performance_curve_b}
     \includegraphics[width=0.45\linewidth]{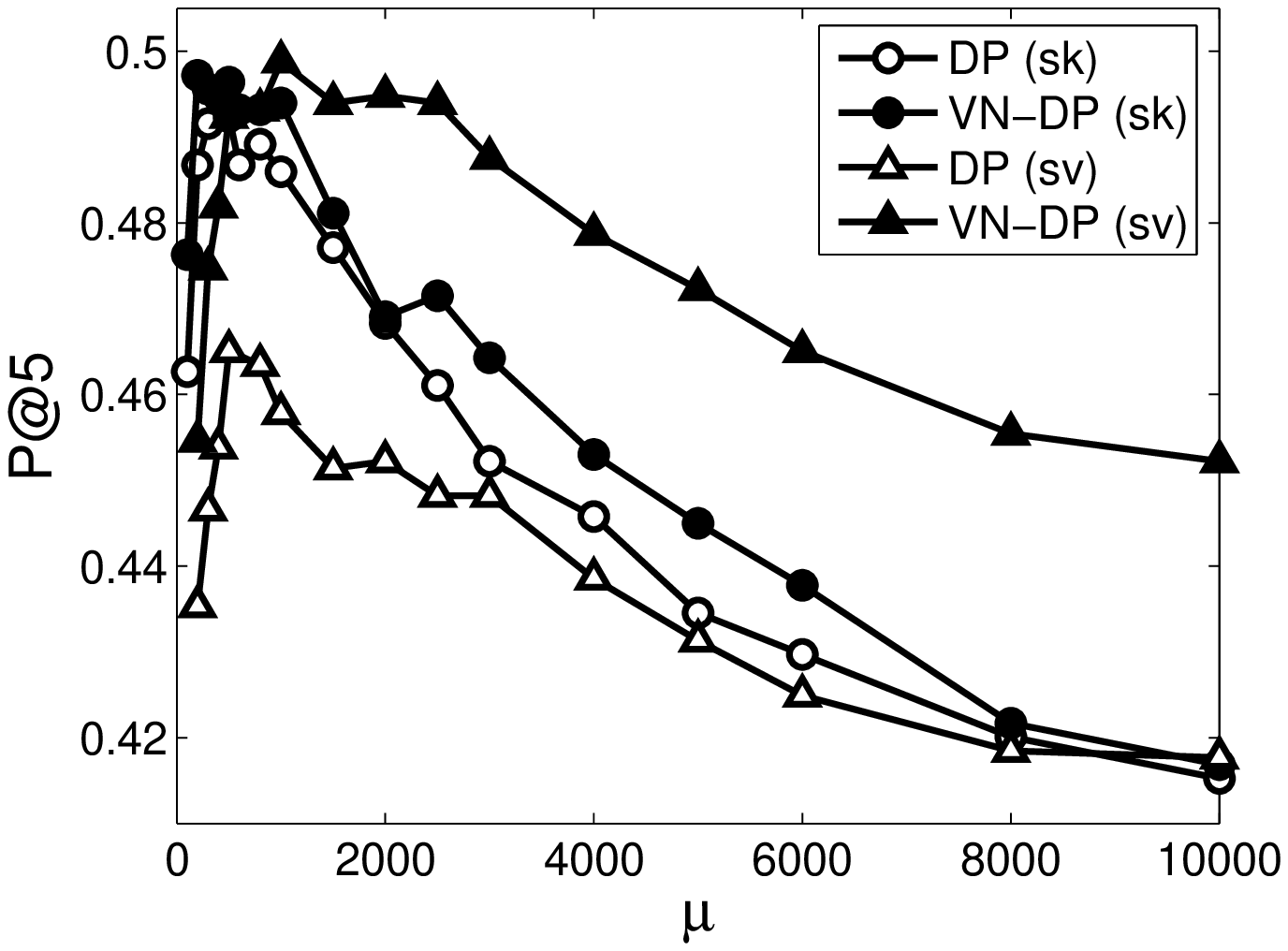}
}
\subfigure[WT10G (MAP)]{
     \label{fig_DP_performance_curve_c}
     \includegraphics[width=0.45\linewidth]{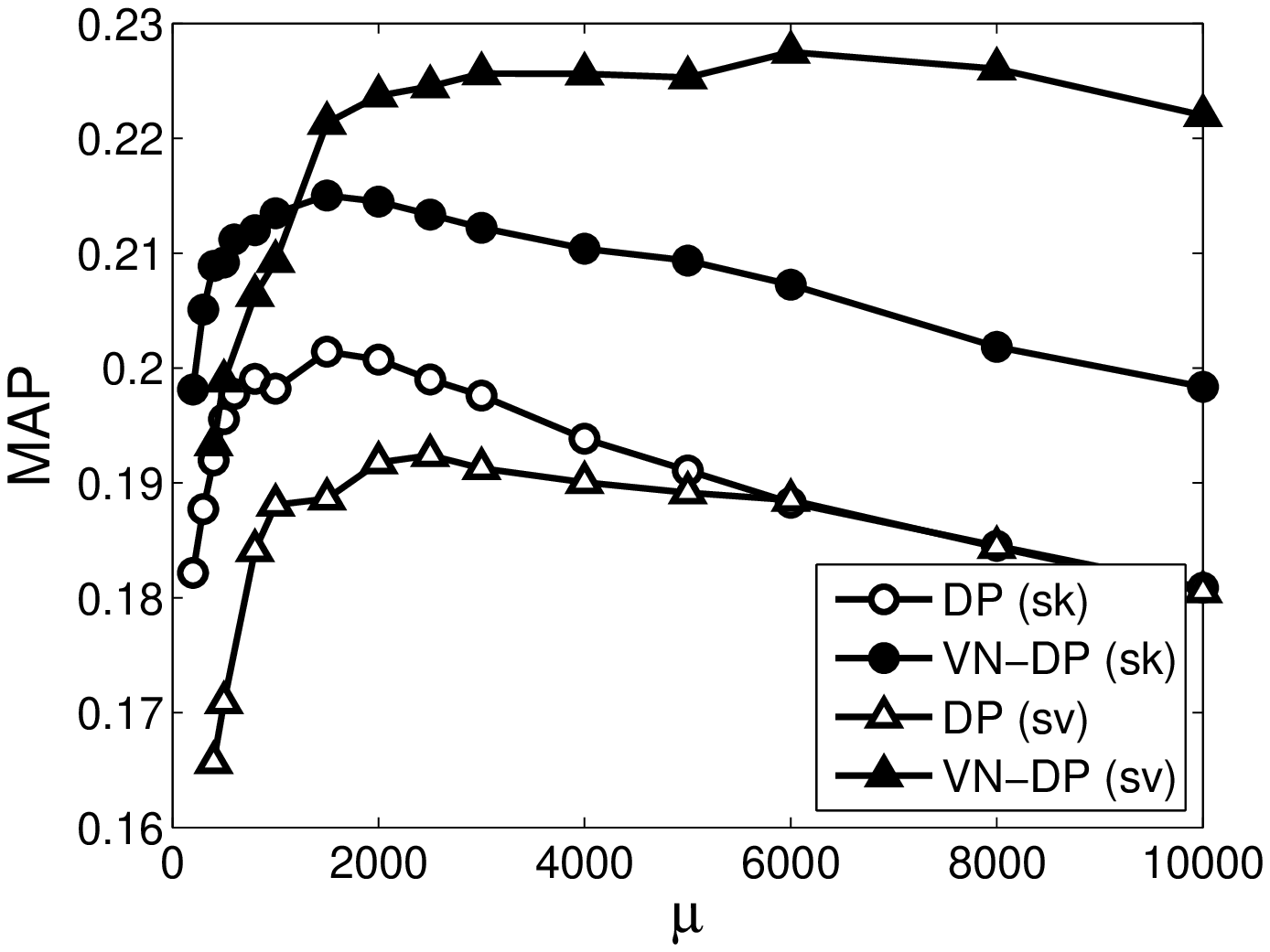}
}
\subfigure[WT10G (P@5)]{
     \label{fig_DP_performance_curve_d}
     \includegraphics[width=0.45\linewidth]{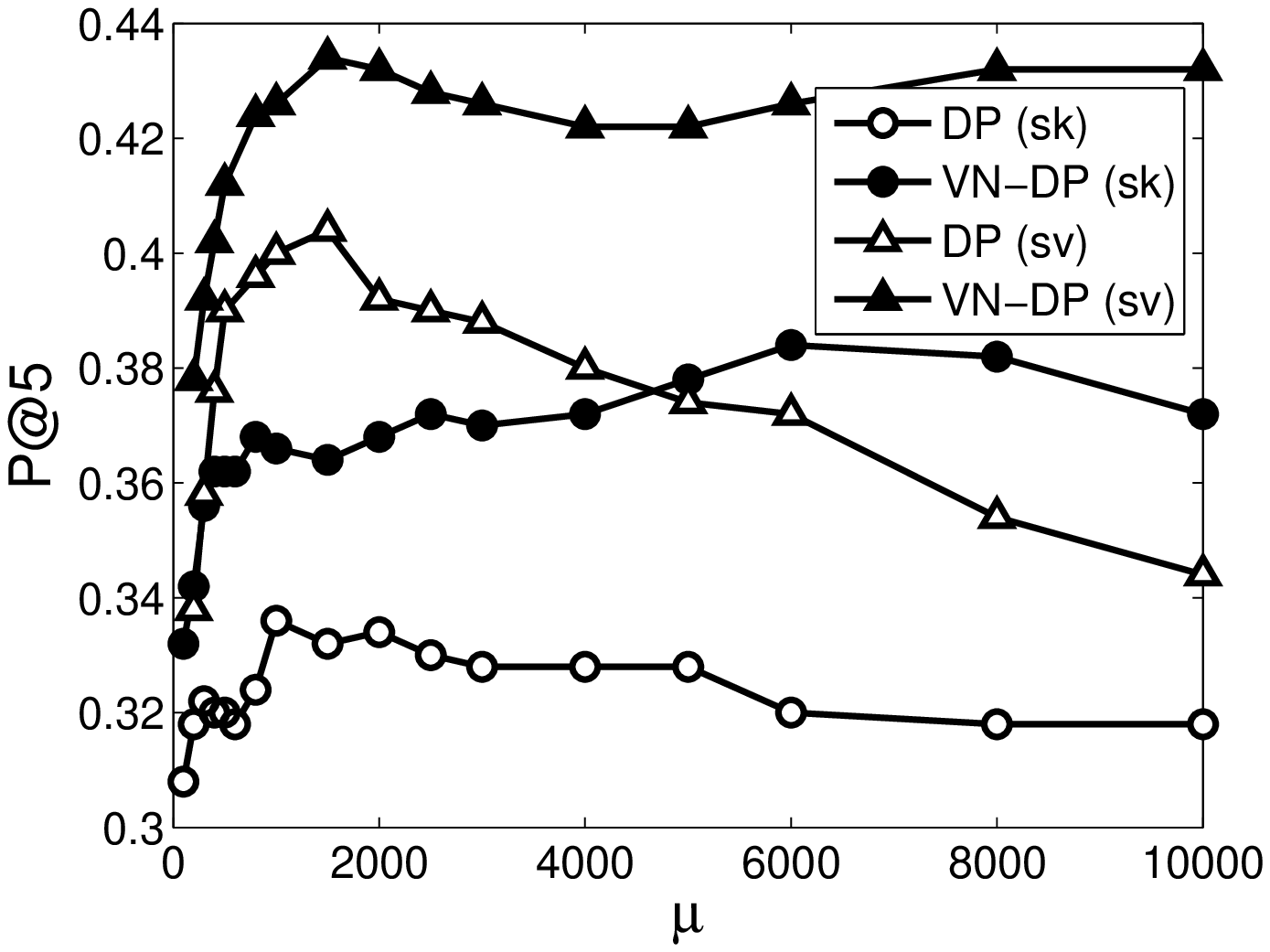}
}
\subfigure[GOV2 (MAP)]{
     \label{fig_DP_performance_curve_e}
     \includegraphics[width=0.45\linewidth]{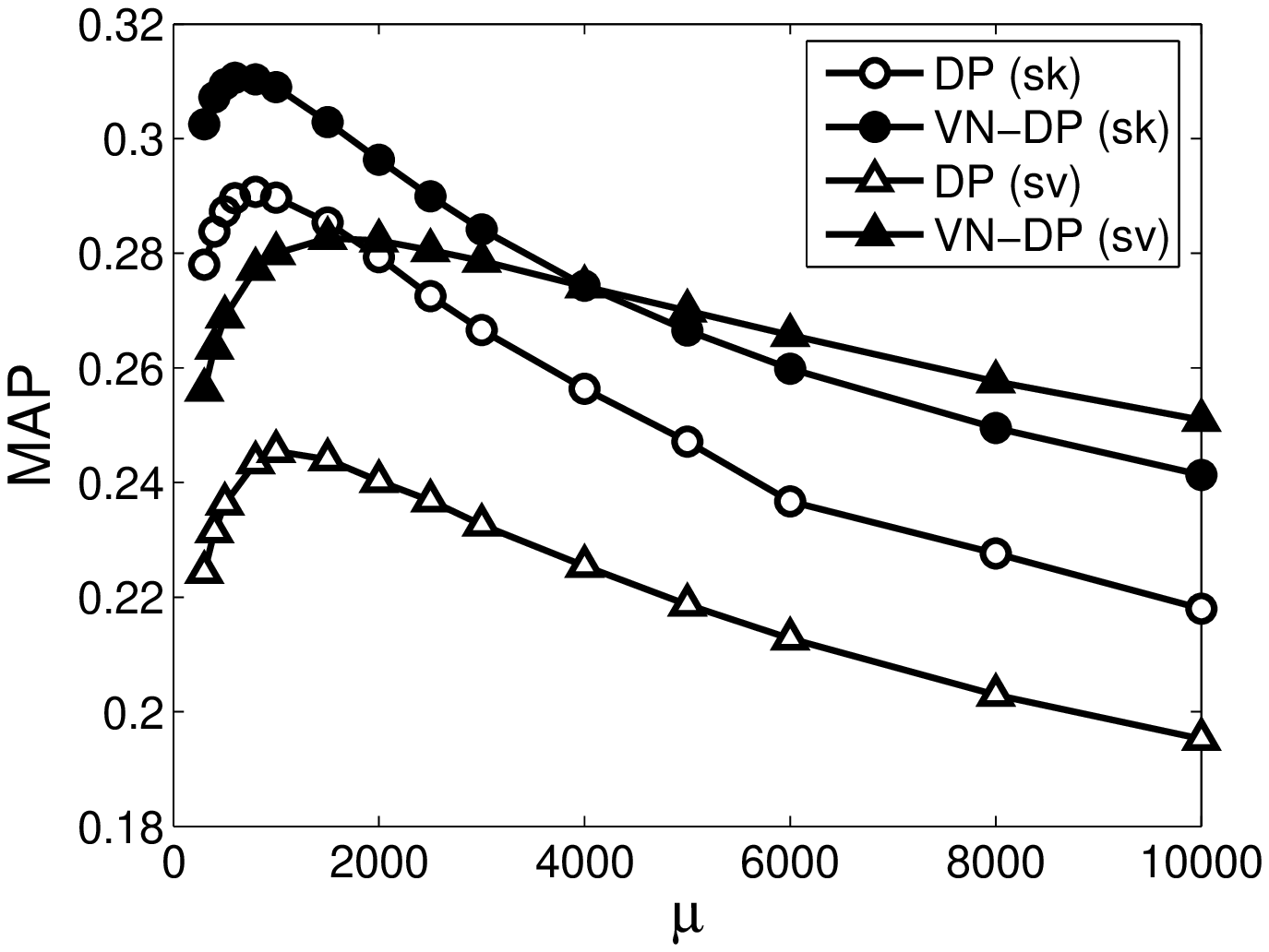}
}
\subfigure[GOV2 (P@5)]{
     \label{fig_DP_performance_curve_f}
     \includegraphics[width=0.45\linewidth]{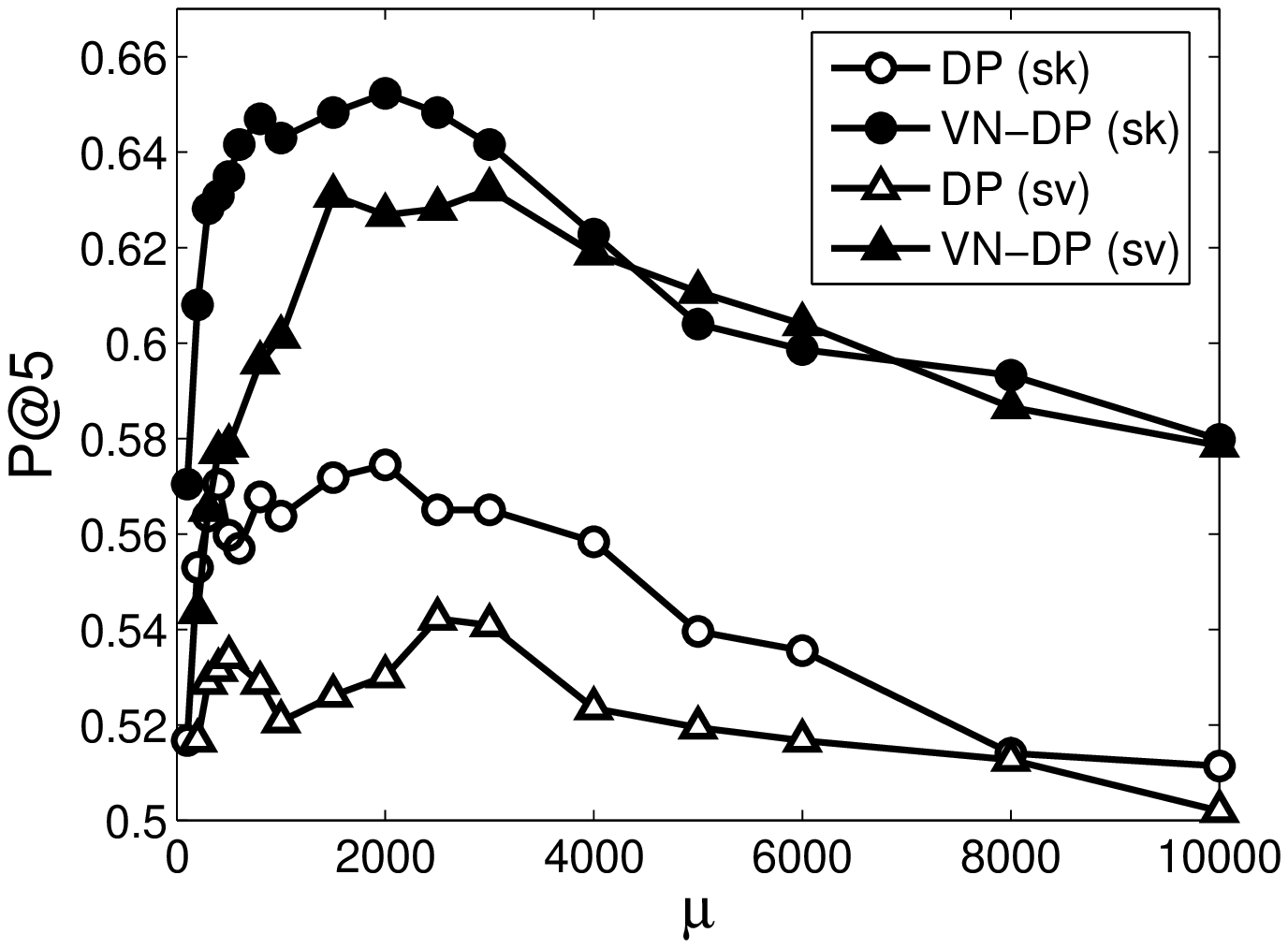}
}
\caption{Performance curves (MAP and P@5) of DP and VN-DP obtained using EntropyPower with varying $\mu$ in ROBUST (top), WT10G (center), and GOV2 (bottom).}
\label{fig_DP_performance_curve}
\end{figure}

\subsection{Okapi vs. VN-Okapi}

Table \ref{tbl_experiment_result_VN_Okapi} show the comparative results (MAP) of Okapi and VN-Okapi under three different scope measures -- UniqLength, EntropyPower, and LengthPower($\beta$).

\begin{table}
\tbl{MAP performance comparison of Okapi and VN-Okapi for three collections, ROBUST, WT10G, and GOV2. The symbols * indicate that a run of the VN method shows statistically significant improvement over the baseline in the t-test at 0.95 confidence level.\label{tbl_experiment_result_VN_Okapi}}{%
\begin{tabular}{||l|c||c|c|c||}\hline
 & Method & \multicolumn{3}{c||}{Okapi (or VN-Okapi)} \\\cline{3-5}
 &  & ROBUST & WT10G & GOV2 \\\hline
\multirow{6}{*}{sk} & baseline & 0.2444 & 0.1946 & 0.2920 \\\cline{2-5}
 & LengthPower(0.5) & 0.2451 & 0.1957 & 0.2897\\\cline{2-5}
 & LengthPower(0.75) & 0.2454 & 0.1994* & 0.2923 \\\cline{2-5}
 & LengthPower(0.9) & 0.2452  & 0.1944 & 0.2923 \\\cline{2-5}
 & UniqLength & \textbf{0.2483}*  & 0.1997 & \textbf{0.3035}* \\\cline{2-5}
 & EntropyPower & 0.2477*  & \textbf{0.2071}*  & 0.3004* \\\hline\hline
\multirow{6}{*}{sv} & baseline & 0.2247 & 0.1853 & 0.2498 \\\cline{2-5}
 & LengthPower(0.5) & 0.2279*  & 0.1806 & 0.2527 \\\cline{2-5}
 & LengthPower(0.75) & 0.2263 & 0.1872 & 0.2530* \\\cline{2-5}
 & LengthPower(0.9) & 0.2271*  & 0.1878 & 0.2529* \\\cline{2-5}
 & UniqLength & 0.2267  & 0.1936 & \textbf{0.2607}*  \\\cline{2-5}
 & EntropyPower & \textbf{0.2303}*  & \textbf{0.1968}* & 0.2599* \\\hline\hline
\multirow{6}{*}{lv} & baseline & 0.2619 & 0.2344 & 0.3012 \\\cline{2-5}
 & LengthPower(0.5) & 0.2647* & 0.2307 & 0.3022 \\\cline{2-5}
 & LengthPower(0.75) & 0.2640* & 0.2341 & 0.3009 \\\cline{2-5}
 & LengthPower(0.9) & 0.2631 & 0.2366 & 0.3018 \\\cline{2-5}
 & UniqLength & \textbf{0.2663}* & 0.2368*  & 0.3063*  \\\cline{2-5}
 & EntropyPower &  0.2659* & \textbf{0.2415}*  & \textbf{0.3074}* \\\hline
\end{tabular}
}
\end{table}
As the results show, VN-Okapi gives improvements; however, the magnitude of these improvements is
smaller than that in the case of VN-DP.  One possible reason for the smaller
improvement is the approximate two-stage normalization carried out in
Okapi, as discussed in Section \ref{section_related_work}. As such, a
form of verbosity normalization is performed by Okapi, to some extent,
using the component $tfn/(tfn + 1)$, which causes VN-Okapi to have
only a limited effect on retrieval performance.

Unlike in the case of DP, there is no significant difference between improvements for
short keyword queries and verbose queries. Therefore, the argument
made for DP wherein H2 is important, particularly for verbose
queries, is much weaker for Okapi. Again, this is because Okapi has its
own component $tfn/(tfn + 1)$ that performs a form of normalization
of verbose documents. As such, excessive preference for a verbose
document is handled to some extent by the original model, even without
our explicit verbosity normalization.

The comparison results for three scope measures are also somewhat different from those of DP. In VN-Okapi, there is no winning scope measure between EntropyPower and UniqLength; in most cases, both  have similar performance.

Despite the limited effects, the improvements obtained by VN-Okapi are
statistically significant, at least for either UniqLength or
EntropyPower, for most of the collections and query types, and thus,
they indicate the merit of our two-stage normalization.

For further comparison, Table \ref{tbl_experiment_result_Okapi_best} presents the best MAPs for Okapi and VN-Okapi and their corresponding parameter values of $b$ and $k_1$ for all three test collections and three query types.
\begin{table}
\tbl{The comparison of the best performance results for Okapi and VN-Okapi using EntropyPower, and the corresponding parameter values ($b$ and $k_1$). The symbols * indicate that a run of the VN-Okapi shows statistically significant improvement over the baseline in the t-test at 0.95 confidence level. \label{tbl_experiment_result_Okapi_best}}{%
\begin{tabular}{||l|c||c|c|c||}\hline
 &  & ROBUST & WT10G & GOV2 \\\hline
\multirow{2}{*}{sk} & Okapi & 0.2454 & 0.2033 & 0.2920 \\\cline{3-5}
& & (0.3, 0.6) & (0.3, 0.6) & (0.01, 0.5)  \\\cline{2-5}
& VN-Okapi & \textbf{0.2482}* & \textbf{0.2107} & \textbf{0.3018}* \\\cline{3-5}
& & (0.1, 0.5)  & (0.05, 0.3)  & (0.1, 0.3) \\\cline{1-5}
\multirow{2}{*}{sv} & Okapi & 0.2267 & 0.1935 & 0.2515 \\\cline{3-5}
& & (0.5, 1.0) & (0.5, 2.0) &  (0.02, 0.6) \\\cline{2-5}
& VN-Okapi & \textbf{0.2303}* & \textbf{0.2001} & \textbf{0.2618}* \\\cline{3-5}
& & (0.3, 0.6)  & (0.3, 1.5) & (0.2, 0.6)  \\\cline{1-5}
\multirow{2}{*}{lv} & Okapi & 0.2637 & 0.2385 &  0.3012 \\\cline{3-5}
& & (0.8, 0.8) & (0.5, 1.5) & (0.03, 0.5) \\\cline{2-5}
 & VN-Okapi & \textbf{0.2676}* & \textbf{0.2482} & \textbf{0.3074}* \\\cline{3-5}
 & & (0.4, 0.6)  & (0.3, 0.8) & (0.3, 0.5) \\\cline{1-5}
\end{tabular}
}
\end{table}

A comparison of the optimal ranges of $b$ across collections for both methods indicates that VN-Okapi tends to be robust without significant differences across collections, whereas Okapi has poor robustness with the optimal values of $b$ being different between GOV2 and other collections. More specifically, for Okapi, the performance surfaces on GOV2 are shifted in the decreasing direction of $b$, relative to those of other collections. As a result, the best performance values of $b$ become much smaller on GOV2 than on other collections for each query type; for short keyword queries, the best value of $b$ is 0.01 on GOV2, and this is smaller than the value of 0.3 on other collections; a similar difference is observed for verbose queries. In contrast, for VN-Okapi, the parameter sensitivity of $b$ on GOV2 is highly similar to that of other collections. The best performance values of $b$ are not different across all collections; the best values of $b$ are commonly between 0.05 and 0.1, for short keyword queries, between 0.2 and 0.3 for short verbose queries, and between 0.3 and 0.4 for long verbose queries.


A comparison of the best performances indicates that VN-Okapi is slightly better than Okapi, in that it highlights the small magnitude of the increase in MAP. Despite its small magnitude, on ROBUST and GOV2, the improvements over Okapi using VN-Okapi are statistically significant for all three types of queries.

\subsection{MRF vs. VN-MRF}
\label{section_expr_mrf}

For evaluating MRF and VN-MRF, because we adopt sequential dependence, a dependency link (undirected link) is inserted only between two \emph{adjacent} query words. Unlike in the case of other query types, for a long verbose query, we do not put a dependency across different topic fields. Thus, no dependency appears between a query word in the title field and a query word in the description or the narrative fields.

Table \ref{tbl_experiment_result_MRF} shows the comparative results of MRF and VN-MRF under three different scope measures, relative to those of DP and VN-DP using EntropyPower.

\begin{table}
\tbl{MAP performance comparison of MRF and VN-MRF on three collections, ROBUST, WT10G, and GOV2, relative to that of DP and VN-DP. The symbols $\alpha$, $\beta$, and $\gamma$ indicate that a run of the VN method shows statistically significant improvement in the t-test at 0.95 confidence level, over DP, VN-DP, and MRF, respectively.\label{tbl_experiment_result_MRF}}{%
\begin{tabular}{||l|c||c|c|c||}\hline
 & Method & \multicolumn{3}{c||}{MRF (or VN-MRF)} \\\cline{3-5}
 &  & ROBUST & WT10G & GOV2 \\\hline
\multirow{8}{*}{sk} & baseline (DP) & 0.2447 & 0.1963 & 0.2907  \\\cline{2-5}
 & baseline (VN-DP) & 0.2481$^{\alpha}$ & 0.2120$^{\alpha}$ & 0.3099$^{\alpha}$  \\\cline{2-5}
 & baseline (MRF) & 0.2545$^{\alpha\gamma}$ & 0.2149$^{\alpha}$ & 0.3095$^{\alpha}$  \\\cline{2-5}
 & LengthPower(0.5) & 0.2506 & 0.2055 & 0.3032  \\\cline{2-5}
 & LengthPower(0.75) & 0.2557$^{\alpha\gamma}$ & 0.2128$^{\alpha}$ & 0.3133$^{\alpha\gamma}$  \\\cline{2-5}
 & LengthPower(0.9) & 0.2545$^{\alpha\gamma}$ & 0.2142$^{\alpha}$ & 0.3125$^{\alpha\gamma}$  \\\cline{2-5}
 & UniqLength & 0.2572$^{\alpha\beta\gamma}$ & 0.2244$^{\alpha\gamma}$ & 0.3270$^{\alpha\beta\gamma}$  \\\cline{2-5}
 & EntropyPower & \textbf{0.2581}$^{\alpha\beta\gamma}$ & \textbf{0.2296}$^{\alpha\beta\gamma}$ &
 \textbf{0.3334}$^{\alpha\beta\gamma}$ \\\hline\hline
\multirow{8}{*}{sv} & baseline (DP) & 0.2260 & 0.1909 & 0.2455 \\\cline{2-5}
 & baseline (VN-DP) & 0.2440$^{\alpha}$ & 0.2196$^{\alpha}$ & 0.2826$^{\alpha\gamma}$ \\\cline{2-5}
 & baseline (MRF) & 0.2416$^{\alpha}$ & 0.2063$^{\alpha}$ & 0.2687$^{\alpha}$ \\\cline{2-5}
 & LengthPower(0.5) & 0.2545$^{\alpha\beta\gamma}$ & 0.2197$^{\alpha}$ & 0.2810$^{\alpha\gamma}$ \\\cline{2-5}
 & LengthPower(0.75) & 0.2507$^{\alpha\beta\gamma}$ & 0.2147$^{\alpha\gamma}$ & 0.2782$^{\alpha\gamma}$  \\\cline{2-5}
 & LengthPower(0.9) & 0.2458$^{\alpha\beta\gamma}$ & 0.2125$^{\alpha\gamma}$ & 0.2739$^{\alpha\gamma}$   \\\cline{2-5}
 & UniqLength & 0.2500$^{\alpha\beta\gamma}$ & 0.2214$^{\alpha\gamma}$ & 0.2879$^{\alpha\gamma}$ \\\cline{2-5}
 & EntropyPower & \textbf{0.2550}$^{\alpha\beta\gamma}$ & \textbf{0.2368}$^{\alpha\beta\gamma}$ & \textbf{0.2975}$^{\alpha\beta\gamma}$  \\\hline\hline
\multirow{8}{*}{lv} & baseline (DP) & 0.2707 & 0.2469 & 0.2864 \\\cline{2-5}
 & baseline (VN-DP) & 0.2799$^{\alpha}$ & 0.2614$^{\alpha}$ & 0.3248$^{\alpha}$ \\\cline{2-5}
 & baseline (MRF) & 0.2813$^{\alpha}$ & 0.2613$^{\alpha}$ & 0.3164$^{\alpha}$ \\\cline{2-5}
 & LengthPower(0.5) & 0.2866$^{\alpha\beta\gamma}$ & 0.2581 & 0.3368$^{\alpha\beta\gamma}$ \\\cline{2-5}
 & LengthPower(0.75) & 0.2883$^{\alpha\beta\gamma}$ & 0.2659$^{\alpha}$ & 0.3280$^{\alpha\gamma}$  \\\cline{2-5}
 & LengthPower(0.9) & 0.2861$^{\alpha\beta\gamma}$ & 0.2617$^{\alpha}$ & 0.3214$^{\alpha\gamma}$ \\\cline{2-5}
 & UniqLength & 0.2895$^{\alpha\beta\gamma}$ & 0.2687$^{\alpha\gamma}$ & 0.3363$^{\alpha\beta\gamma}$ \\\cline{2-5}
 & EntropyPower & \textbf{0.2927}$^{\alpha\beta\gamma}$ & \textbf{0.2754}$^{\alpha\beta\gamma}$ & \textbf{0.3481}$^{\alpha\beta\gamma}$  \\\hline
\end{tabular}
}
\end{table}

It is clearly seen that MRF is always better than DP, with all of the performance improvements being statistically significant. This precisely reproduces the comparison results reported by the existing works on MRF \cite{metzler05mrf}. Note that the improved performance using MRF is further enhanced by VN-MRF with the application of the two-stage normalization, and additional improvements are statistically significant improvements. In particular, either on UniqLength or EntropyPower, VN-MRF is always better than MRF for all test collections and all query types, with all improvements being statistically significant.

Interestingly, VN-DP alone without exploiting the term dependency is nearly comparable to MRF, often even showing better performances. Again, the performance of VN-DP is further increased by VN-MRF along with the utilization of the term dependency, and the additional improvements are statistically significant in most cases, at least using either EntropyPower or UniqLength. Therefore, this result strongly implies that both effects resulting from the term dependency and the two-stage normalization are slightly co-related, thus facilitating such incremental increase by their combined utilization.

Another interesting result is that the performance difference of VN-MRF across test collections and query types shows a highly similar tendency to that of VN-DP. First, both VN methods (VN-MRF and VN-DP) are more effective, especially on the heterogeneous web collections (WT10G and GOV2) than on ROBUST. Second, on EntropyPower, both VN methods show larger improvements for verbose queries than for keyword queries -- the only exception is found in VN-MRF for long verbose query on WT10G, where the improvement is slightly smaller than that for short keyword queries. Third, on LengthPower, both VN methods often show improvements over their original methods, and they are more effective for verbose queries than for keyword queries.

The similarity between the two VN methods is understandable considering the fact that the underlying retrieval function in MRF is basically the same as that of DP -- DP and MRF commonly employ the smoothed document model of Eq. (\ref{eq_DirichletVN_smoothed_model}) for scoring a document.

Table \ref{tbl_experiment_result_MRF_P5} shows the performances of P@5 of VN-MRF, in comparison to those of MRF, using the values of the MAP-optimized parameters. The results for P@5 are largely similar to those for MAP, as seen in Table \ref{tbl_experiment_result_MRF}. In many cases, the MRF's performance of P@5 is better than that of DP, often with statistically significant improvements. Further, the performance of MRF is increased by VN-MRF with two-stage normalization, at least using either UniqLength or EntropyPower, and often with statistically significant improvements. In the particular case using EntropyPower, VN-MRF yields statistically significant improvements over MRF on WT10G and GOV2 for all short keyword queries and on ROBUST and GOV2 for some verbose queries. This result implies that in many cases, VN-MRF's significant improvement in MAP results from the increased performance of P@5. VN-DP alone shows performance similar to that of MRF. Again, the precision of VN-DP is slightly increased by VN-MRF exploiting the term dependency, although usually not to a statistically significant degree, unlike the results in MAP.

\begin{table}
\tbl{Comparison of performance of P@5 of MRF and VN-MRF for three collections, ROBUST, WT10G, and GOV2, relative to that of DP and VN-DP. The symbols $\alpha$, $\beta$, and $\gamma$ indicate that a run of the VN method shows statistically significant improvement in the t-test at 0.95 confidence level, over DP, VN-DP, and MRF, respectively.\label{tbl_experiment_result_MRF_P5}}{%
\begin{tabular}{||l|c||c|c|c||}\hline
 & Method & \multicolumn{3}{c||}{MRF (or VN-MRF)} \\\cline{3-5}
 &  & ROBUST & WT10G & GOV2 \\\hline
\multirow{8}{*}{sk} & baseline (DP) & 0.4924 & 0.3120 & 0.5678  \\\cline{2-5}
 & baseline (VN-DP) & 0.4972 & 0.3640$^{\alpha}$ & 0.6416 \\\cline{2-5}
 & baseline (MRF) & 0.5036 & 0.3580$^{\alpha}$ & 0.6121 \\\cline{2-5}
 & LengthPower(0.5) & 0.4859 & 0.3540$^{\alpha}$ & 0.5664 \\\cline{2-5}
 & LengthPower(0.75) & 0.4916 & 0.3500$^{\alpha}$ & 0.6054$^{\alpha}$ \\\cline{2-5}
 & LengthPower(0.9) & 0.5004 & 0.3540$^{\alpha}$ & 0.6121$^{\alpha}$ \\\cline{2-5}
 & UniqLength & \textbf{0.5068} & 0.3660$^{\alpha}$ & 0.6470$^{\alpha\gamma}$ \\\cline{2-5}
 & EntropyPower & 0.5012 & \textbf{0.3840}$^{\alpha\beta\gamma}$ & \textbf{0.6685}$^{\alpha\gamma}$ \\\hline\hline
\multirow{8}{*}{sv} & baseline (DP) & 0.4466 & 0.3880 & 0.5208 \\\cline{2-5}
 & baseline (VN-DP) & 0.4932$^{\alpha}$ & 0.4300$^{\alpha}$ & 0.6309 \\\cline{2-5}
 & baseline (MRF) & 0.4876$^{\alpha}$ & 0.4240$^{\alpha}$ & 0.5839 \\\cline{2-5}
 & LengthPower(0.5) & 0.4916$^{\alpha}$ & 0.4140 & 0.5544 \\\cline{2-5}
 & LengthPower(0.75) & 0.4972$^{\alpha}$ & 0.4160  & 0.5785$^{\alpha}$ \\\cline{2-5}
 & LengthPower(0.9) & 0.4892$^{\alpha}$ & 0.4120  & 0.5812$^{\alpha}$ \\\cline{2-5}
 & UniqLength & 0.4940$^{\alpha}$ & 0.4320$^{\alpha}$ & 0.5919$^{\alpha\gamma}$ \\\cline{2-5}
 & EntropyPower & \textbf{0.5044}$^{\alpha\gamma}$ & \textbf{0.4400}$^{\alpha}$ & \textbf{0.6376}$^{\alpha\gamma}$  \\\hline\hline
\multirow{8}{*}{lv} & baseline (DP) & 0.5414 & 0.4460 & 0.6228 \\\cline{2-5}
 & baseline (VN-DP) & 0.5631$^{\alpha}$ & 0.4700 & 0.6644$^{\alpha}$ \\\cline{2-5}
 & baseline (MRF) & 0.5598$^{\alpha}$ & 0.4700$^{\alpha}$ & 0.6550$^{\alpha}$ \\\cline{2-5}
 & LengthPower(0.5) & 0.5582 & 0.4680 & 0.6336 \\\cline{2-5}
 & LengthPower(0.75) & 0.5639$^{\alpha}$ & 0.4720$^{\alpha}$ & 0.6376 \\\cline{2-5}
 & LengthPower(0.9) & 0.5655$^{\alpha}$ & 0.4580 & 0.6456 \\\cline{2-5}
 & UniqLength & 0.5751$^{\alpha\gamma}$ & \textbf{0.4760}$^{\alpha}$ & 0.6671$^{\alpha}$ \\\cline{2-5}
 & EntropyPower & \textbf{0.5799}$^{\alpha\beta\gamma}$ & 0.4640 & \textbf{0.6886}$^{\alpha\gamma}$\\\hline
\end{tabular}
}
\end{table}

For further comparison, Figure \ref{fig_MRF_performance_curve} shows the performance curves of MRF and VN-MRF, plotted by varying $\mu$, for short keyword and verbose queries -- EntropyPower is used as the scope measure, and MAP and P@5 are used as the evaluation measures. Again, there is a great degree of similarity between the comparison results of VN-MRF and MRF and those of VN-DP and DP -- for P@5 curves, VN-MRF is always better than the original method, except for only a few parameter values of $\mu$s in ROBUST. The shapes of the performance curves of P@5 are quite similar for both VN-MRF and MRF. The optimal ranges of $\mu$ are also close, as in the case of VN-DP and DP.

\begin{figure}[ht]
\centering
\subfigure[ROBUST (MAP)]{
     \label{fig_MRF_performance_curve_a}
     \includegraphics[width=0.45\linewidth]{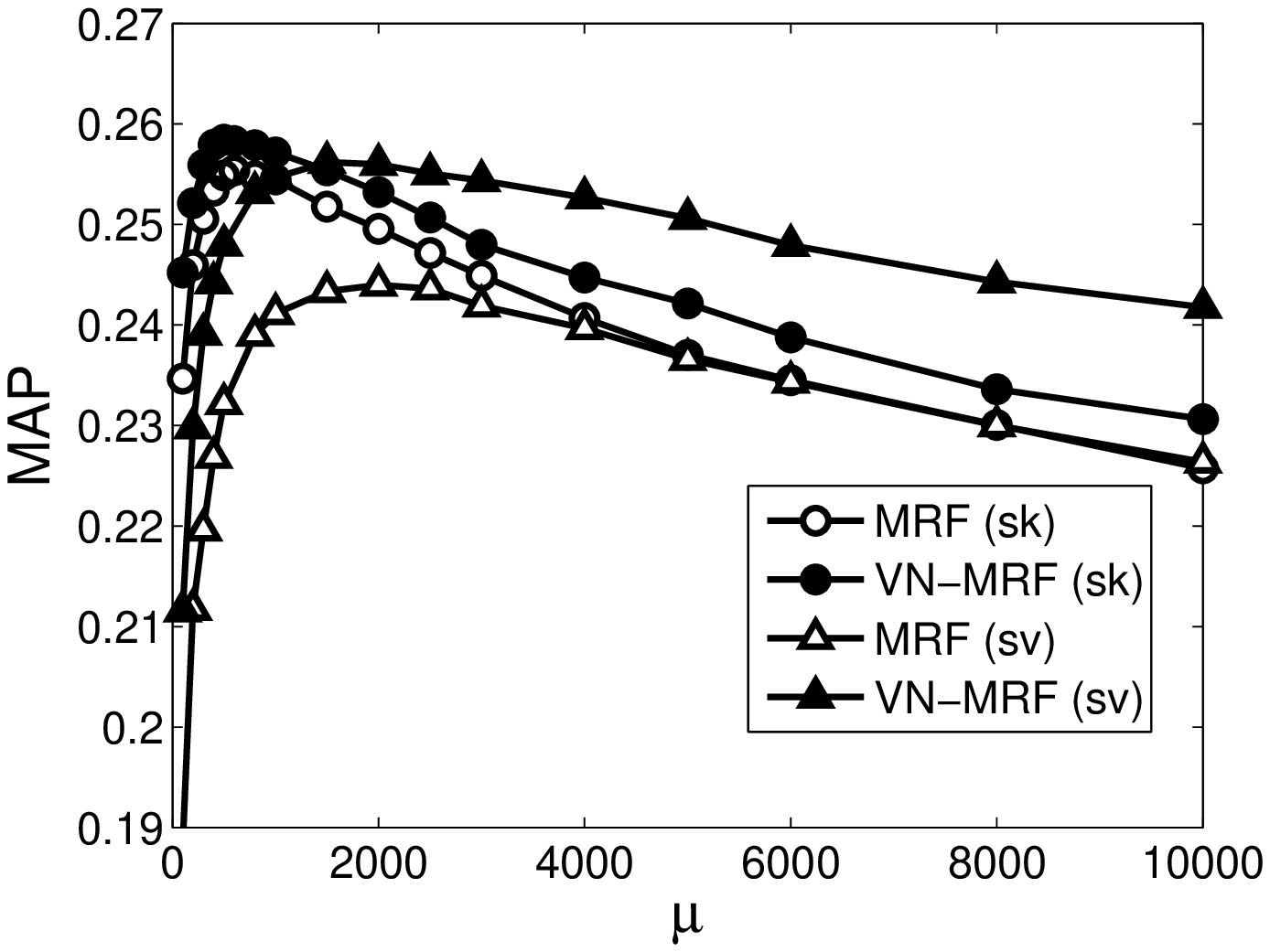}
}
\subfigure[ROBUST (P@5)]{
     \label{fig_MRF_performance_curve_b}
     \includegraphics[width=0.45\linewidth]{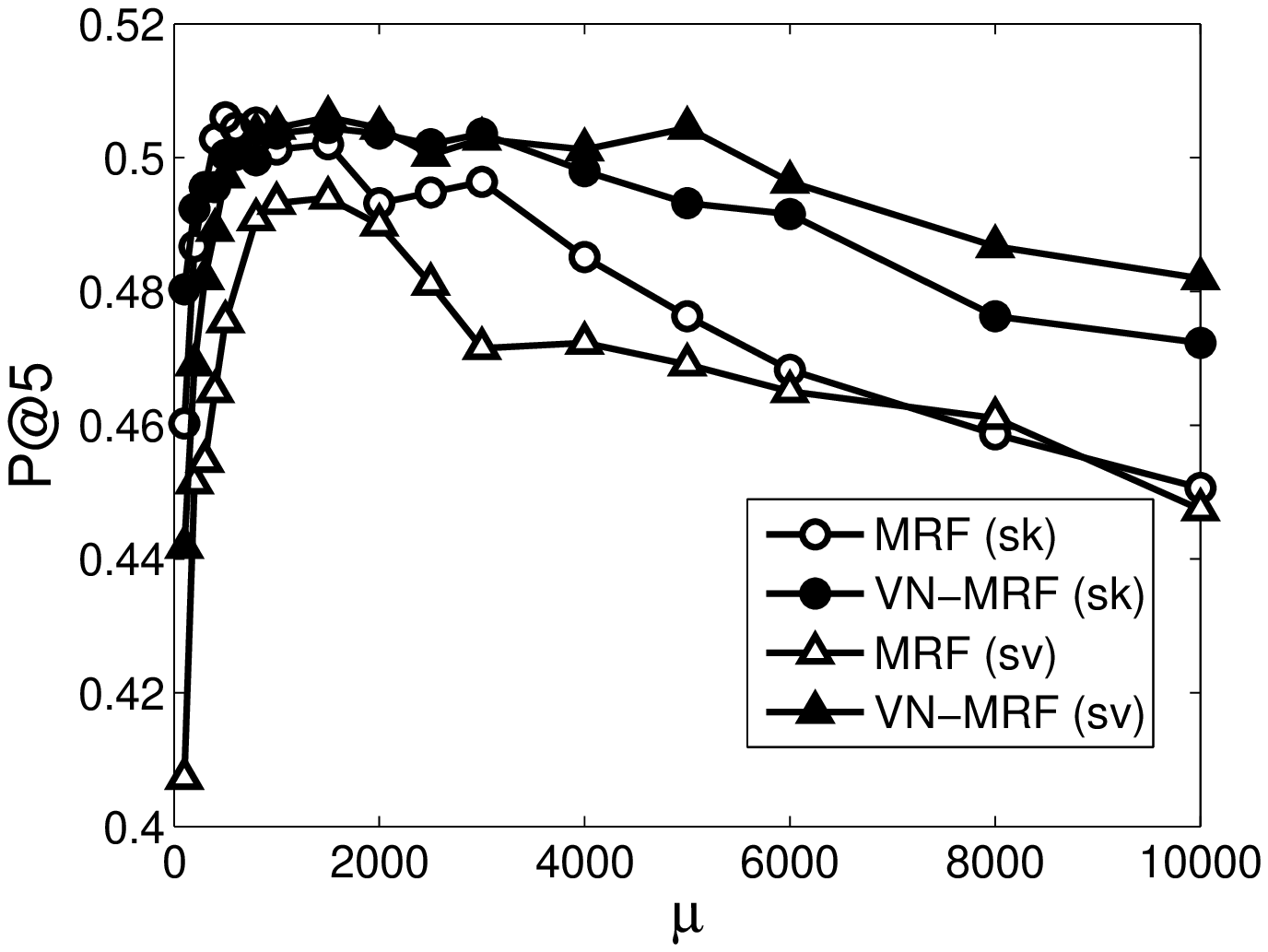}
}
\subfigure[WT10G (MAP)]{
     \label{fig_MRF_performance_curve_c}
     \includegraphics[width=0.45\linewidth]{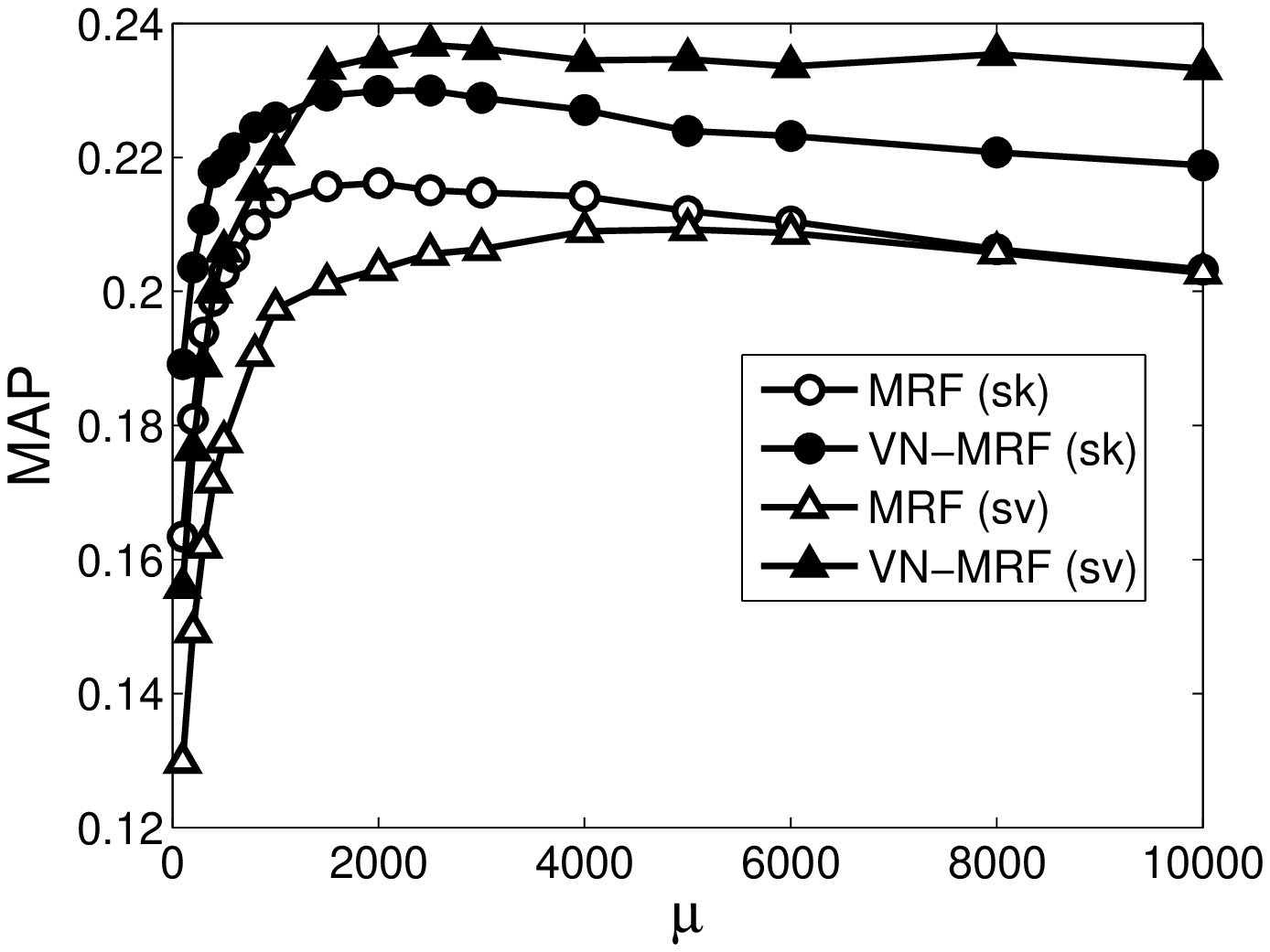}
}
\subfigure[WT10G (P@5)]{
     \label{fig_MRF_performance_curve_d}
     \includegraphics[width=0.45\linewidth]{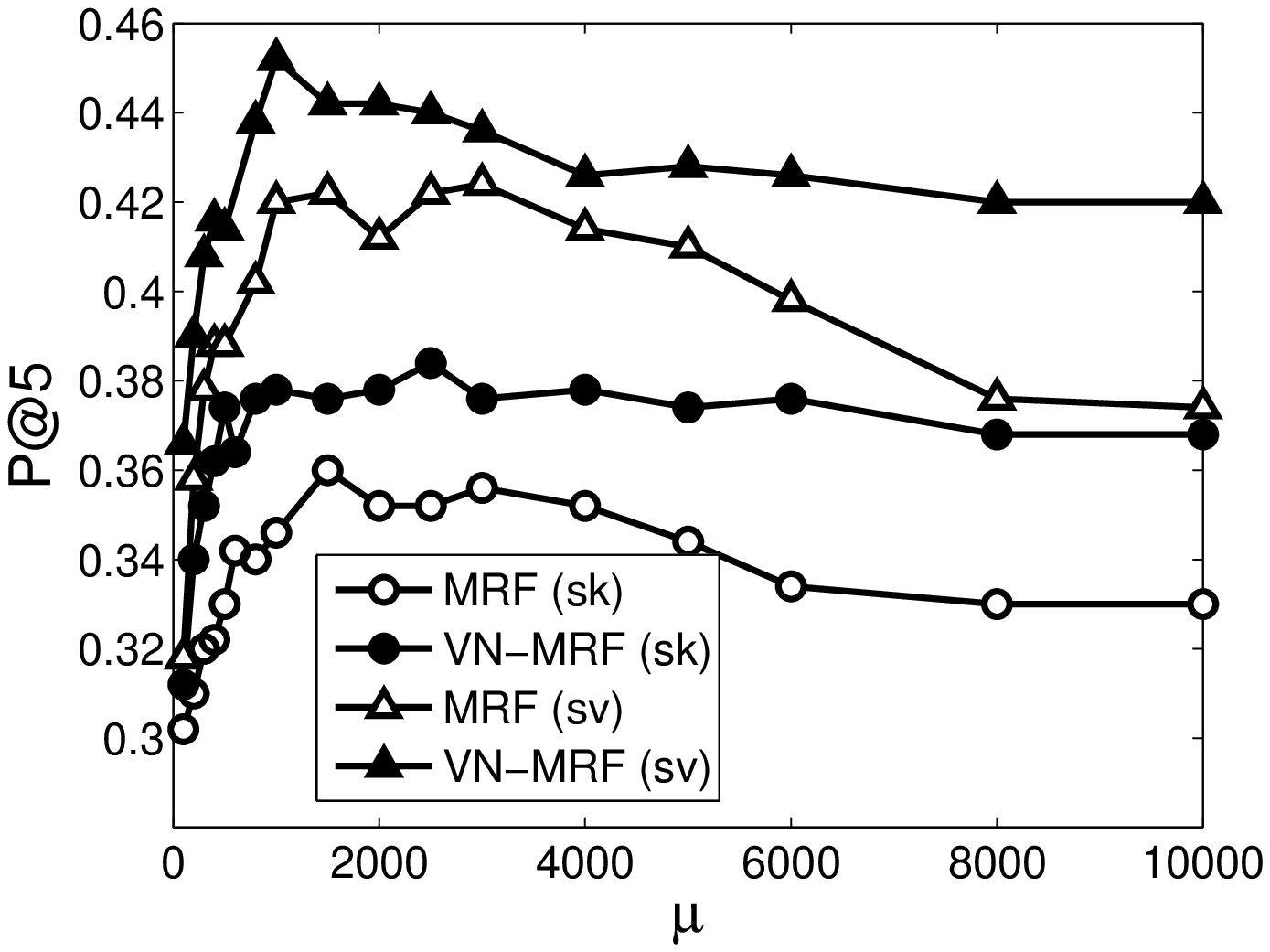}
}
\subfigure[GOV2 (MAP)]{
     \label{fig_MRF_performance_curve_e}
     \includegraphics[width=0.45\linewidth]{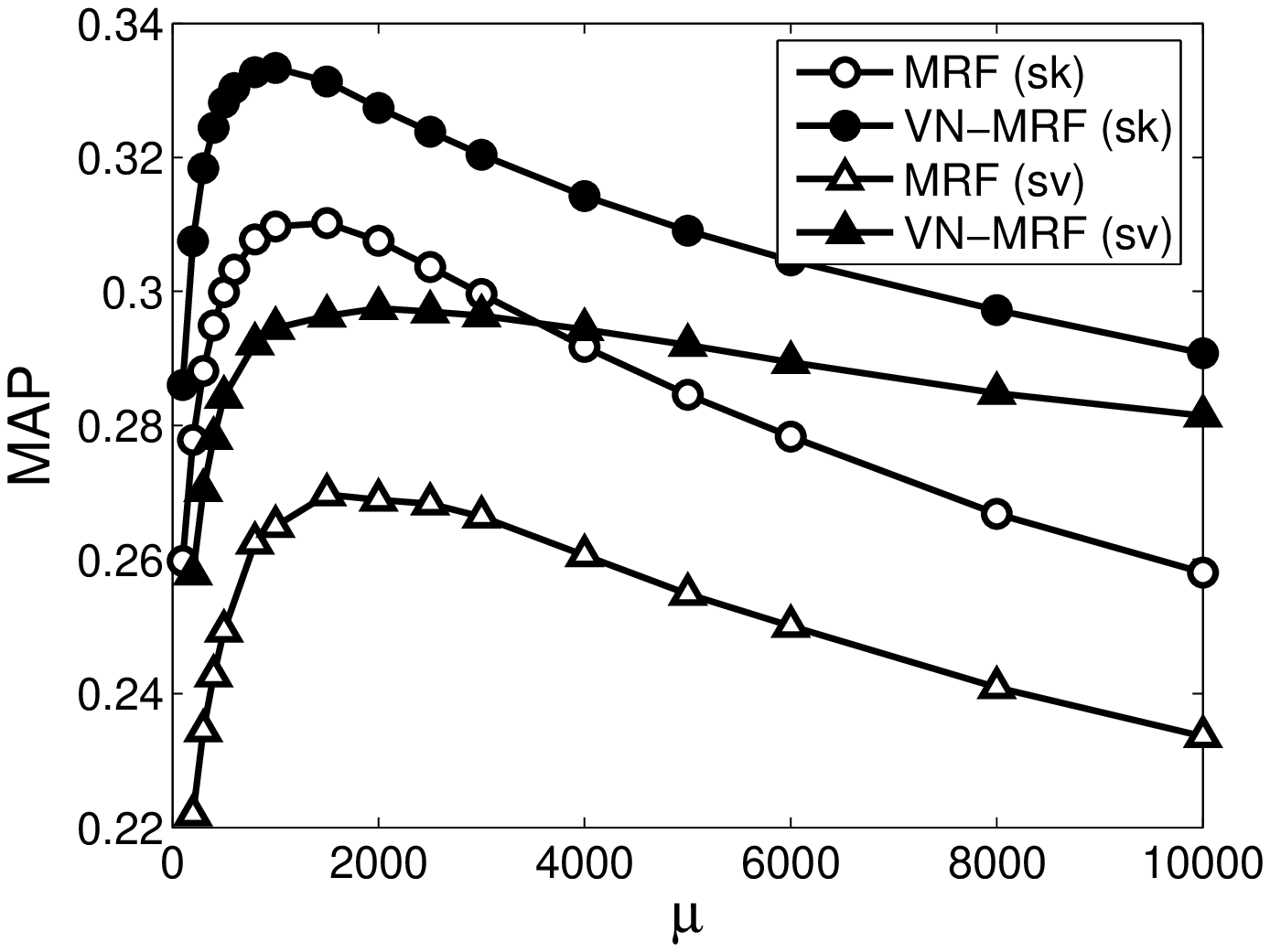}
}
\subfigure[GOV2 (P@5)]{
     \label{fig_MRF_performance_curve_f}
     \includegraphics[width=0.45\linewidth]{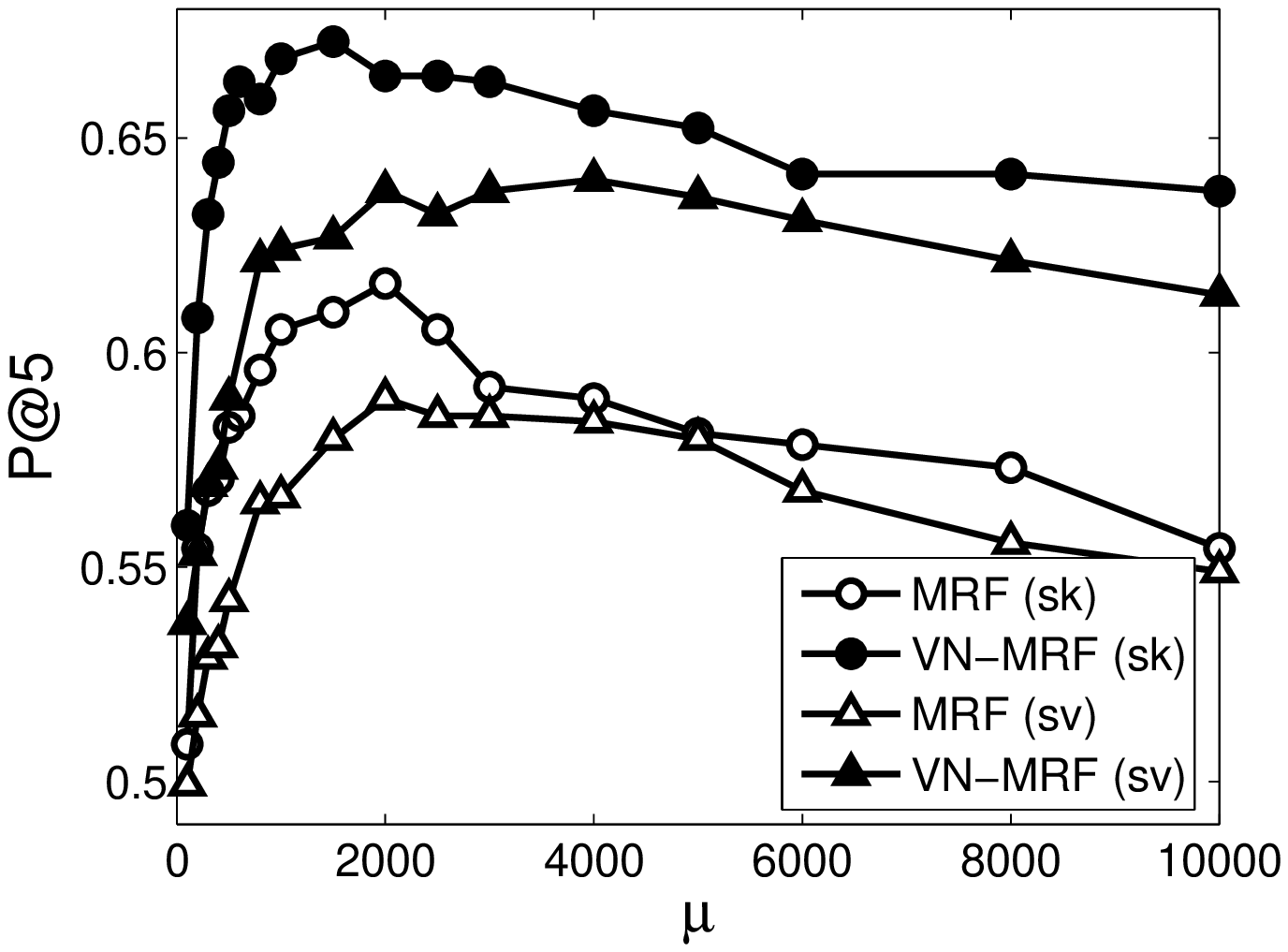}
}
\caption{Performance curves (MAP and P@5) of MRF and VN-MRF obtained using EntropyPower with varying $\mu$ on ROBUST (top), WT10G (center), and GOV2 (bottom).}
\label{fig_MRF_performance_curve}
\end{figure}

\section{APPLICATION TO LOWER BOUNDING TERM FREQUENCY NORMALIZATION}
\label{section_application_lower_bounding_term_freq_normalization}

\subsection{Lower-Bounded Retrieval Models}
As discussed in related works, \cite{lv09b} recently proposed the use of lower-bounding term frequency normalization in avoiding over-penalization of very long documents. Their experimental results showed that lower-bounded retrieval models lead to significant improvements in comparison with baseline models. An interesting issue is whether our proposed two-stage normalization can further improve these lower-bounded models. To this end, we chose DP and VN-DP as retrieval models, and compared their lower-bounded models with their VN models.

Two lower-bounded models for DP and VN-DP are presented in the following. First, a lower-bounded model for DP can be formulated as follows:
\begin{eqnarray}
\sum_{ w \in q \cap d} c(w,q)\left[ \ln\left( 1 + \frac{c(w,d)}{\mu P(w|C)} \right)  + \ln\left( 1 + \frac{\delta}{\mu P(w|C)} \right) \right]
 + |q| \ln\left( \frac{\mu}{|d| + \mu} \right)
 \label{eq_LB_Dirichlet_formula}
\end{eqnarray}
which is called \textbf{DP+}. In Eq. (\ref{eq_LB_Dirichlet_formula}), $\delta$ is a pseudo term frequency value that controls the scale of the lower bound, which was introduced by \cite{lv11b}.

Second, a lower-bounded model for VN-DP can be formulated by straightforwardly applying the general normalization approach of \cite{lv11b}\footnote{According to the notation of \cite{lv11b}, $F(c(w,\phi(d)),|\phi(d)|,td(w))$ corresponds to $\ln\left( \frac{\mu}{s(d)+ \mu} + \frac{c(w,\phi(d)) }{(s(d)+ \mu) P(w|C)} \right)$ }.
\begin{eqnarray}
\sum_{ w \in q \cap d} && c(w,q) \left[ \ln\left( 1 + \frac{c(w,d)}{\mu P(w|C)} \frac{s(d)}{|d|} \right)  + \ln\left( 1 + \frac{\delta}{\mu P(w|C)} \right) \right] \nonumber \\
&& + |q| \ln\left( \frac{\mu}{s(d) + \mu} \right)
 \label{eq_LB_DirichletVN_formula}
\end{eqnarray}
which is called \textbf{VN-DP+}.

Similarly, we can derive a lower-bounded Okapi (\textbf{Okapi+}) and a lower-bounded VN-Okapi (\textbf{VN-Okapi+}). In the original BM25 retrieval formula, Okapi+ uses $tf_{BM25+}(w,d)$ (i.e., $tf_{BM25}(w,d)$ + $\delta$) for the term frequency component, and VN-Okapi+ uses $tf_{BM25}(w, \phi(d))$ (i.e., $tf_{BM25}(w,\phi(d))$ + $\delta$) which are given by
\begin{eqnarray}
tf_{BM25+}(w,d) = \frac{(k_1 + 1) c(w,d)}{k_1 \left( \left( 1-b\right) + b|d| /avgl\right) +c(w,d)} + \delta
 \label{eq_LB_OkapiTF_formula}
\end{eqnarray}
\begin{eqnarray}
tf_{BM25+}(w,\phi(d)) = \frac{(k_1 + 1) c(w,d)}{k_1 |d| \left( \left( (1-b)/s(d)\right) + b /avgs\right) +c(w,d)} + \delta
 \label{eq_LB_OkapiVN_TF_formula}
\end{eqnarray}

\subsection{Experiment Results}
For parameter tuning of the lower-bounded models, we follow the setting in the work of \cite{lv09b}; For DP+ and VN-DP+, we search $\delta$ over the space between 0 and 0.15, with increments of 0.01.
For Okapi+ and VN-Okapi+, we search $\delta$ over space between 0.0 and 1.5, with the increment of 0.1. The search spaces for other retrieval parameters are the same as those used in previous sections.

Table \ref{tbl_experiment_result_LB_VN_DP} shows the MAP performances of DP+ and VN-DP+, as compared to DP and VN-DP. As shown in Table \ref{tbl_experiment_result_LB_VN_DP}, DP+ often exhibits non-trivial improvements over DP, especially for short verbose queries, reaffirming the results reported in \cite{lv11b} that DP+ shows greater effectiveness for short verbose queries than for keyword queries, which shows a different result from that achieved by \cite{lv11b}; our experiment shows a statistically significant improvement when using DP+ over DP for only sv queries in the ROBUST collection. Unlike DP+, VN-DP+, the lower-bounded model over VN-DP, does not show greater effectiveness for short verbose queries than for other types of queries. This may be because VN-DP already shows a significant improvement over DP for short verbose queries, and further improvement is therefore less likely. Nevertheless, VN-DP+ continues to further increase the performances of VN-DP, with improvements being statistically significant for short keyword queries in GOV2 and long verbose queries in ROBUST.


\begin{table}
\tbl{MAP performance comparison of DP and VN-DP on three collections ROBUST, WT10G, and GOV2, and three different query types sk, sv, and lv.
EntropyPower is used for the scope measure in VN-DP and VN-DP+. Symbols $\alpha$, $\beta$, and $\gamma$ indicate that a run of the VN method (or a lower-bounded method) shows a statistically significant improvement over DP, DP+, VN-DP, respectively, in the t-test at 0.95 confidence level.\label{tbl_experiment_result_LB_VN_DP}}{%
\begin{tabular}{||l|c||c|c|c||}\hline
 & Method & \multicolumn{3}{c||}{DP+ (or VN-DP+)} \\\cline{3-5}
 &  & ROBUST & WT10G & GOV2 \\\hline
\multirow{4}{*}{sk} & DP & 0.2447 & 0.1963 & 0.2907 \\\cline{2-5}
 & DP+ & 0.2447 & 0.1957 & 0.2922 \\\cline{2-5}
 & VN-DP & \textbf{0.2481}$^{\alpha\beta}$ & \textbf{0.2120}$^{\alpha\beta}$ & 0.3099$^{\alpha\beta}$ \\\cline{2-5}
 & VN-DP+ & 0.2476$^{\alpha\beta}$ & 0.2112$^{\alpha\beta}$ & \textbf{0.3141}$^{\alpha\beta\gamma}$ \\\hline
\multirow{4}{*}{sv}  & DP & 0.2260 & 0.1909 & 0.2455 \\\cline{2-5}
 & DP+ & 0.2337$^{\alpha}$ & 0.1969 & 0.2453 \\\cline{2-5}
 & VN-DP & 0.2440$^{\alpha\beta}$ & 0.2196$^{\alpha\beta}$ & \textbf{0.2826}$^{\alpha\beta}$ \\\cline{2-5}
 & VN-DP+ & \textbf{0.2461}$^{\alpha\beta}$ & \textbf{0.2215}$^{\alpha\beta}$ & 0.2819$^{\alpha\beta}$ \\\hline
\multirow{4}{*}{lv} & DP & 0.2707 & 0.2469 & 0.2864 \\\cline{2-5}
 & DP+ & 0.2766 & 0.2442 & 0.2863 \\\cline{2-5}
 & VN-DP & 0.2799$^{\alpha}$ & \textbf{0.2614}$^{\alpha\beta}$ & 0.3248$^{\alpha\beta}$ \\\cline{2-5}
 & VN-DP+ & \textbf{0.2858}$^{\alpha\beta\gamma}$ & 0.2603$^{\alpha\beta}$ & \textbf{0.3248}$^{\alpha\beta}$ \\\hline
\end{tabular}
}
\end{table}

Importantly, on comparing VN-DP with DP+, we can see that DP+ does not reach the performance of VN-DP. For almost all runs (except for lv in ROBUST), the improvements gained by VN-DP over DP are mostly larger than those made by DP+ over DP, and in most cases are statistically significant. Furthermore, VN-DP leads mostly to statistically significant improvements over DP+ for almost all runs. These results clearly demonstrate that the improvement from the VN model over DP is not redundant to the effects from the existing lower-bounding normalization, and leads to a significant improvement even against lower-bounded models, which are stronger baselines. Overall, our experimental results indicate that two-stage normalization significantly improves lower-bounded models for almost all runs for three different collections.


We now consider the comparison between lower-bounded models for Okapi and VN-Okapi and their original models. Table \ref{tbl_experiment_result_LB_VN_Okapi} lists the MAP performances of Okapi+ and VN-Okapi+, as compared to those of Okapi and VN-Okapi. Again, results similar to those presented in Table \ref{tbl_experiment_result_LB_VN_DP} are obtained, although the improvements by the VN models over lower-bounded models are not larger than the case of DP; the lower-bounded models are effective in improving baseline models, without reaching the performance of VN-Okapi. The improvements gained by VN-Okapi over Okapi are mostly larger than those made by Okapi+ over Okapi.
Although the improvements of VN-Okapi over Okapi+ are not statistically significant in most cases, VN-Okapi+ leads to statistically significant improvements over Okapi+ for almost all runs.



\begin{table}
\tbl{MAP performance comparison of Okapi and VN-Okapi on three collections ROBUST, WT10G, and GOV2, and three different query types sk, sv, and lv.
EntropyPower is used for the scope measure in VN-Okapi and VN-Okapi+. Symbols $\alpha$, $\beta$, and $\gamma$ indicate that a run of the VN method (or a lower-bounded method) shows a statistically significant improvement over Okapi, Okapi+, VN-Okapi, respectively, in the t-test at 0.95 confidence level.\label{tbl_experiment_result_LB_VN_Okapi}}{%
\begin{tabular}{||l|c||c|c|c||}\hline
 & Method & \multicolumn{3}{c||}{Okapi+ (or VN-Okapi+)} \\\cline{3-5}
 &  & ROBUST & WT10G & GOV2 \\\hline
\multirow{4}{*}{sk} & Okapi & 0.2444 & 0.1946 & 0.2920 \\\cline{2-5}
 & Okapi+ & 0.2457$^{\alpha}$ & 0.2039$^{\alpha}$ & 0.2969$^{\alpha}$ \\\cline{2-5}
 & VN-Okapi & 0.2477$^{\alpha}$ & 0.2071$^{\alpha}$ & 0.3004$^{\alpha}$ \\\cline{2-5}
 & VN-Okapi+ & \textbf{0.2477}$^{\alpha}$ & \textbf{0.2085}$^{\alpha}$ & \textbf{0.3100}$^{\alpha\beta\gamma}$ \\\hline
\multirow{4}{*}{sv}  & Okapi & 0.2247 & 0.1884 & 0.2498 \\\cline{2-5}
 & Okapi+ & 0.2279$^{\alpha}$ & 0.1900 & 0.2573$^{\alpha}$ \\\cline{2-5}
 & VN-Okapi & 0.2303$^{\alpha}$ & 0.1968$^{\alpha}$ & 0.2599$^{\alpha}$\\\cline{2-5}
 & VN-Okapi+ & \textbf{0.2311}$^{\alpha\beta}$ & \textbf{0.2023}$^{\alpha\gamma}$ & \textbf{0.2658}$^{\alpha\beta\gamma}$ \\\hline
\multirow{4}{*}{lv} & Okapi & 0.2619 & 0.2314 & 0.3012 \\\cline{2-5}
 & Okapi+ & 0.2640$^{\alpha}$ & 0.2320 & 0.3059$^{\alpha}$ \\\cline{2-5}
 & VN-Okapi & \textbf{0.2659}$^{\alpha}$ & \textbf{0.2415}$^{\alpha\beta}$ & 0.3074$^{\alpha}$ \\\cline{2-5}
 & VN-Okapi+ & 0.2658$^{\alpha}$ & 0.2390$^{\alpha\beta}$ & \textbf{0.3094}$^{\alpha\beta}$  \\\hline
\end{tabular}
}
\end{table}



For further comparison, we present the performances of original, VN, and lower-bounded models  with respect to standard topic sets of TREC in three test collections, named TREC6, TREC7, TREC8, ROBUST03, ROBUST04, TREC9, TREC10, TREC2004, TREC2005, and TREC2006.
Table \ref{tbl_standard_topic_set_TREC} presents the basic information on the standard topic sets of TREC.

First, Table \ref{tbl_test_performance_DP} shows the MAP performances between DP+ and VN-DP+, as compared to DP and VN-DP on standard TREC topic sets.
As shown in Table  \ref{tbl_test_performance_DP}, VN-DP or VN-DP+ show further improvements over DP and DP+ for almost all standard topic sets
and they are statistically significant for more than half of all cases. In particular, VN-DP+ shows the best performance
for almost all runs. Their improvements over DP are statistically significant (except for standard topic sets of sk queries in ROBUST and lv queries in WT10G) and are larger than the improvements of VN-DP or DP+ over DP.
Comparing VN-DP to DP+, more runs showed improvements of statistical significance on VN-DP over DP than on DP+ over DP.



\begin{table}
\tbl{Standard topic sets of TREC, their corresponding collection names, and their training topic sets.\label{tbl_standard_topic_set_TREC}}{%
\begin{tabular}{||l|c|c|c||}\hline
Topic set id & Query ids & Collection & Training topic sets\\\hline
TREC6 & Q301-Q350 & \multirow{5}{*}{ROBUST} & TREC7,TREC8,ROBUST03,ROBUST04 \\\cline{1-2}\cline{4-4}
TREC7 & Q351-Q400 &  & TREC6,TREC8,ROBUST03,ROBUST04 \\\cline{1-2}\cline{4-4}
TREC8 & Q401-Q450 &  & TREC6,TREC7,ROBUST03,ROBUST04 \\\cline{1-2}\cline{4-4}
ROBUST03 & Q601-Q650 & & TREC6,TREC7,TREC8,ROBUST04 \\\cline{1-2}\cline{4-4}
ROBUST04 & Q651-Q700 & & TREC6,TREC7,TREC8,ROBUST03 \\\hline
TREC9 & Q451-Q500 & \multirow{2}{*}{WT10G} & TREC10 \\\cline{1-2}\cline{4-4}
TREC10 & Q501-Q550 &  & TREC9 \\\hline
TREC2004 & Q701-Q750 & \multirow{2}{*}{GOV2} & TREC2005,TREC2006 \\\cline{1-2}\cline{4-4}
TREC2005 & Q751-Q800 &  & TREC2004,TREC2006 \\\cline{1-2}\cline{4-4}
TREC2006 & Q801-Q850 &  & TREC2004,TREC2005 \\\hline
\end{tabular}
}
\end{table}




\begin{table}
\tbl{MAP performance comparison of DP, DP+, VN-DP, and VN-DP+ on standard topic sets in TREC and three different query types sk, sv, and lv.
EntropyPower is used for the scope measure in VN-DP and VN-DP+. Symbols $\alpha$, $\beta$, and $\gamma$ indicate that a run of the VN method (or a lower-bounded method) shows a statistically significant improvement over DP,
DP+, VN-DP, respectively, in the t-test at 0.95 confidence level.\label{tbl_test_performance_DP}}{%
\begin{tabular}{||l|l||p{1.3cm}|p{1.3cm}|p{1.5cm}|p{1.6cm}||}\hline
 &  & DP & DP+ & VN-DP & VN-DP+ \\\hline
\multirow{10}{*}{sk} & TREC6 & 0.2465 & 0.2471 & \textbf{0.2483} & 0.2479 \\\cline{2-6}
& TREC7 & 0.1733 & 0.1733 & \textbf{0.1785} & 0.1783 \\\cline{2-6}
& TREC8 & 0.2410 & 0.2415 & 0.2425 & \textbf{0.2425} \\\cline{2-6}
& ROBUST03 & 0.2756 & 0.2755 & \textbf{0.2818}$^{\alpha\beta}$ & 0.2812$^{\alpha\beta}$  \\\cline{2-6}
& ROBUST04 & 0.2879 & 0.2871 & \textbf{0.2903} & 0.2892 \\\cline{2-6}
& TREC9 & 0.1985 & 0.1997 & 0.2063 & \textbf{0.2068} \\\cline{2-6}
& TREC10 & 0.1942 & 0.1917 & \textbf{0.2177}$^{\alpha\beta}$ & 0.2156$^{\alpha\beta}$ \\\cline{2-6}
& TREC2004 & 0.2597 & 0.2605 & 0.2790$^{\alpha\beta}$ & \textbf{0.2842}$^{\alpha\beta}$  \\\cline{2-6}
& TREC2005 & 0.3114 & 0.3130 & 0.3297$^{\alpha\beta}$ & \textbf{0.3336}$^{\alpha\beta}$  \\\cline{2-6}
& TREC2006 & 0.3005 & 0.3026 & 0.3205$^{\alpha\beta}$ & \textbf{0.3228}$^{\alpha\beta}$ \\\hline
\multirow{10}{*}{sv} & TREC6 & 0.1751 & 0.1898 & 0.1980$^{\alpha}$ & \textbf{0.2018}$^{\alpha\beta}$ \\\cline{2-6}
& TREC7 & 0.1698 & 0.1768$^{\alpha}$ & 0.1887$^{\alpha\beta}$ & \textbf{0.1904}$^{\alpha\beta}$   \\\cline{2-6}
& TREC8 & 0.2145 & 0.2194 & 0.2301 & \textbf{0.2306}$^{\alpha\beta}$  \\\cline{2-6}
& ROBUST03 & 0.2912 & 0.2996 & 0.3171$^{\alpha\beta}$ & \textbf{0.3182}$^{\alpha\beta}$  \\\cline{2-6}
& ROBUST04 & 0.2806 & 0.2840 & 0.2872 & \textbf{0.2906}$^{\alpha}$  \\\cline{2-6}
& TREC9 & 0.1996 & 0.2094$^{\alpha}$  & 0.2299 & \textbf{0.2325}$^{\alpha}$  \\\cline{2-6}
& TREC10 & 0.1822 & 0.1844 & 0.2093$^{\alpha\beta}$  & \textbf{0.2105}$^{\alpha\beta}$   \\\cline{2-6}
& TREC2004 & 0.2163 & 0.2154 & \textbf{0.2498}$^{\alpha\beta}$ & 0.2475$^{\alpha\beta}$ \\\cline{2-6}
& TREC2005 & 0.2524 & 0.2524 & 0.2860$^{\alpha\beta}$ & \textbf{0.2860}$^{\alpha\beta}$  \\\cline{2-6}
& TREC2006 & 0.2671 & 0.2676 & 0.3114 $^{\alpha\beta}$ & \textbf{0.3114}$^{\alpha\beta}$   \\\hline
\multirow{10}{*}{lv} & TREC6 		& 0.2627 & 0.2792$^{\alpha}$  & 0.2713$^{\alpha}$ & \textbf{0.2904}$^{\alpha\beta}$ \\\cline{2-6}
& TREC7 		& 0.2152 & 0.2254$^{\alpha}$ & 0.2219$^{\alpha}$ & \textbf{0.2296}$^{\alpha\beta}$  \\\cline{2-6}
& TREC8 		& 0.2421 & 0.2511$^{\alpha}$  & 0.2539$^{\alpha}$ & \textbf{0.2619}$^{\alpha\beta\gamma}$   \\\cline{2-6}
& ROBUST03 	& 0.3289 & 0.3252 & 0.3391$^{\alpha\beta}$ & \textbf{0.3401}$^{\alpha\beta}$  \\\cline{2-6}
& ROBUST04 	& 0.3051 & 0.3063 & \textbf{0.3106} & 0.3074 \\\cline{2-6}
& TREC9 		& 0.2550 & 0.2550 & \textbf{0.2675} & 0.2675 \\\cline{2-6}
& TREC10 		& 0.2388 & 0.2334 & \textbf{0.2553}$^{\alpha\beta}$ & 0.2530 \\\cline{2-6}
& TREC2004 	& 0.2650 & 0.2651 & 0.2943$^{\alpha\beta}$  & \textbf{0.2943}$^{\alpha\beta}$ \\\cline{2-6}
& TREC2005 	& 0.2875 & 0.2870 & 0.3232$^{\alpha\beta}$  & \textbf{0.3232}$^{\alpha\beta}$  \\\cline{2-6}
& TREC2006 	& 0.3062 & 0.3062 & 0.3563$^{\alpha\beta}$ & \textbf{0.3563}$^{\alpha\beta}$ \\\hline
\end{tabular}
}
\end{table}


Turning to the comparison between lower-bounded and VN models for Okapi,
Table \ref{tbl_test_performance_Okapi} shows the MAP performances between Okapi+ and VN-Okapi+,
as compared to Okapi and VN-Okapi on standard TREC topic sets. Again, VN-Okapi+ shows the best performance for almost all runs,
and their improvements over Okapi are larger than those made by VN-Okapi or Okapi+ over Okapi, being statistically significant for most cases.

\begin{table}
\tbl{MAP performance comparison of Okapi, Okapi+, VN-Okapi, and VN-Okapi+ on standard topic sets in TREC and three different query types sk, sv, and lv.
EntropyPower is used for the scope measure in VN-Okapi and VN-Okapi+. Symbols $\alpha$, $\beta$, and $\gamma$ indicate that a run of the VN method (or a lower-bounded method) shows a statistically significant improvement over Okapi,
Okapi+, VN-Okapi, respectively, in the t-test at 0.95 confidence level. \label{tbl_test_performance_Okapi}}{%
\begin{tabular}{||l|l||p{1.3cm}|p{1.3cm}|p{1.5cm}|p{1.6cm}||}\hline
 &  &  Okapi & Okapi+ & VN-Okapi &  VN-Okapi+ \\\hline
\multirow{10}{*}{sk} & TREC6 & 0.2471 & 0.2479 & 0.2492 & \textbf{0.2498} \\\cline{2-6}
& TREC7 & 0.1734 & 0.1754 & 0.1758 & \textbf{0.1773}$^{\alpha}$ \\\cline{2-6}
& TREC8 & 0.2413 & 0.2415 & \textbf{0.2445} & 0.2416 \\\cline{2-6}
& ROBUST03 & 0.2760 & 0.2786 & 0.2826$^{\alpha}$ & \textbf{0.2833}$^{\alpha}$ \\\cline{2-6}
& ROBUST04 & 0.2848 & 0.2861 & 0.2872 & \textbf{0.2874} \\\cline{2-6}
& TREC9 & 0.1946 & 0.2077$^{\alpha}$ & 0.2050 & \textbf{0.2089} \\\cline{2-6}
& TREC10 & 0.1946 & 0.2001 & \textbf{0.2092}$^{\alpha}$ & 0.2081 \\\cline{2-6}
& TREC2004 & 0.2562 & 0.2606 & 0.2659 & \textbf{0.2766}$^{\alpha\beta\gamma}$  \\\cline{2-6}
& TREC2005 & 0.3222 & 0.3297$^{\alpha}$ & 0.3233 & \textbf{0.3354}$^{\alpha\gamma}$ \\\cline{2-6}
& TREC2006 & 0.2969 & 0.2996 & 0.3113$^{\alpha}$ & \textbf{0.3174}$^{\alpha\beta}$  \\\hline
\multirow{10}{*}{sk} & TREC6 & 0.1628 & 0.1663 & 0.1678 & \textbf{0.1678} \\\cline{2-6}
& TREC7 & 0.1603 & 0.1642$^{\alpha}$  & 0.1653$^{\alpha}$ & \textbf{0.1702}$^{\alpha\beta\gamma}$ \\\cline{2-6}
& TREC8 & 0.2111 & 0.2144 & 0.2154 & \textbf{0.2172}$^{\alpha}$ \\\cline{2-6}
& ROBUST03 & 0.3148 & 0.3145 & \textbf{0.3196}$^{\alpha}$ & 0.3159 \\\cline{2-6}
& ROBUST04 & 0.2753 & 0.2809$^{\alpha}$  & 0.2847 &  \textbf{0.2847}$^{\alpha}$ \\\cline{2-6}
& TREC9 &  0.1958 & 0.1970 & 0.2055 & \textbf{0.2117}$^{\alpha\gamma}$ \\\cline{2-6}
& TREC10 & 0.1809 & 0.1831 & 0.1880 & \textbf{0.1930}$^{\alpha}$ \\\cline{2-6}
& TREC2004 & 0.2240 & 0.2302 & 0.2327$^{\alpha}$ & \textbf{0.2401}$^{\alpha\beta\gamma}$ \\\cline{2-6}
& TREC2005 & 0.2540 & 0.2682$^{\alpha}$ & 0.2597 & \textbf{0.2714}$^{\alpha\gamma}$ \\\cline{2-6}
& TREC2006 &0.2707 & 0.2731 & \textbf{0.2866}$^{\alpha}$  & 0.2855$^{\alpha\beta}$ \\\hline
\multirow{10}{*}{lv} & TREC6 	 &  0.2365 & 0.2360 & 0.2366 & \textbf{0.2374}	\\\cline{2-6}
& TREC7 	 &  0.2077 & \textbf{0.2104}$^{\alpha}$ & 0.2093 & 0.2093		\\\cline{2-6}
& TREC8 	 &  0.2420 & 0.2434 & 0.2448 & \textbf{0.2451}   \\\cline{2-6}
& ROBUST03 &  0.3287 & 0.3331 & 0.3378 & \textbf{0.3381}$^{\alpha}$    \\\cline{2-6}
& ROBUST04 &  0.2954 & 0.2980 & \textbf{0.3016}$^{\alpha}$ & 0.3000$^{\alpha}$    \\\cline{2-6}
& TREC9 	 &  0.2248 & 0.2281$^{\alpha}$ & \textbf{0.2413}$^{\alpha\beta}$ & 0.2362$^{\alpha\beta}$	\\\cline{2-6}
& TREC10 	 &  0.2379 & 0.2359 & 0.2418 & \textbf{0.2418}   \\\cline{2-6}
& TREC2004 &  0.2695 & 0.2710 & 0.2743 & \textbf{0.2746}      \\\cline{2-6}
& TREC2005 &  0.3056 & 0.3146$^{\alpha}$ & 0.3083 & \textbf{0.3164}$^{\alpha}$   \\\cline{2-6}
& TREC2006 &  0.3280 & 0.3315 & \textbf{0.3389}$^{\alpha}$& 0.3365$^{\alpha\beta}$       \\\hline
\end{tabular}
}
\end{table}

Thus, the main results of Tables \ref{tbl_test_performance_DP} and \ref{tbl_test_performance_Okapi} are largely consistent with those reported
in Tables \ref{tbl_experiment_result_LB_VN_DP} and \ref{tbl_experiment_result_LB_VN_Okapi}, respectively. The lower bounding models do not reach the performance of VN models;
the improvements gained by VN models over the baseline are mostly larger than those made by lower bounding models over the baseline;
the further improvements even against lower bounding models are made by lower bounding VN models (VN-DP+ or VN-Okapi+)
\footnote{
However, compared to the previous experiments, the statistical significance for VN models is
weakly supported on some sets of standard queries for both the DP and Okapi cases.
The result of the statistical significance is also fairly observed in lower bounding models
where the improvements are not statistically significant on some of the standard TREC topic sets.
We believe that the reason is the lack of evidence for  by which to judge significance.
The number of queries used in standard topics is 50, which is often not sufficient to convincingly decide significance,
especially when the improvement is marginal. Consider, for example, the case of Okapi and sk queries in ROBUST (shown in Table \ref{tbl_test_performance_Okapi}).
When using only 50 queries in standard TREC, the improvements made by VN-Okapi and Okapi+
over the baseline are mostly not of statistical significance. However, when using 250 queries in ROBUST,
the improvements turn out to be statistically significant as shown in Table \ref{tbl_experiment_result_LB_VN_Okapi}.
Thus, the results show that a larger number of queries might be necessary for the statistical
significance test when the improvements are marginal.
}.

As a consequence, the overall results shown in Table \ref{tbl_experiment_result_LB_VN_DP}, \ref{tbl_experiment_result_LB_VN_Okapi}, \ref{tbl_test_performance_DP}, and \ref{tbl_test_performance_Okapi}
consistently indicate that the improvement resulting from the application of two-stage normalization
is not fully replaceable by adopting lower-bounding term frequency normalization, and vice versa.
This result is intuitive, as two normalizations aim at different deficiencies of the existing
normalization method: lower-bounding term frequency normalization aims at avoiding
the penalization of \emph{very} long documents, while two-stage normalization
aims at avoiding \emph{insufficient} penalization of verbose documents and \emph{excessive} penalization
of long documents. The resultant solutions for these different goals also differ:
lower-bounding term frequency normalization does not need to decompose
the document length into additional factors, but rather enforces
the addition of scope gaps between document scores when a query term appears and disappears in a document.
Two-stage normalization decomposes a document length into verbosity and scope factors, penalizing verbose and broad documents \emph{separately}.

In conclusion, given the set of results described throughout this section, we can state the following. The proposed two-stage normalization is clearly effective for further improving the existing retrieval model; it is not limitedly applicable to improving the baseline retrieval model and can also be extended even to the improved methods that uses  the term dependency and the lower-bounding term frequency normalization. From the comparative axiomatic analysis results, we conclude that the normalization heuristics H1 and H2 should necessarily be applied for scoring a document.

\section{CONCLUSION}
\label{section_conclusion}

In this paper, we argue that a normalization function should use different
penalizations for verbosity and scope, and we propose the use of
two-stage normalization. Our main contributions over and above those
of previous works that formulated ranking functions belonging to
two-stage normalization \cite{singhal96,na08} are as follows: 1) We
generalize two-stage normalization such that it can
be applied to any retrieval model.  2) We perform comparative
axiomatic analysis and capture the exact retrieval heuristics resulting
from two-stage normalization and its difference from the original
method. The results of experiments on three models -- DP, Okapi, and MRF
-- consistently show that two-stage normalization is promising.

Of course, considerable work needs to be done in the future. Although
two-stage normalization is effective in the case of DP, Okapi, and MRF, we
still need more evidence of its effectiveness for other retrieval
models. Thus, an obvious future work is to explore the application of
two-stage normalization to the pivoted vector space model
\cite{singhal96}, DFR \cite{amati02,amati02b}, a more recently developed
information model \cite{clinchant10}, a parameterized query expansion \cite{bendersky11},
a \emph{term-specific} adaptation of normalization parameter \cite{lv11c}, or a learning-to-rank framework \cite{liu09,li11}.
Another research direction is strengthening the axiomatic framework by generalizing the current
retrieval constraints such that they can effectively cover the
conjectured retrieval heuristics derived in this paper.  A more
challenging future research direction is to develop a new retrieval
model in an innovative manner such that it includes the verbosity, scope, and
document length as retrieval parameters.

\bibliographystyle{ACM-Reference-Format-Journals}
\bibliography{shna_verbosity_compact}  


\section*{Appendix A: Definition of Verbosity}
For full definition of verbosity, suppose that $M$ is the total number of \emph{topics} in a collection, and $s(d)$ is the number of topics mentioned in $d$ (or the \emph{expected} number of topics in $d$).
Here, we assume that the topic is \emph{countable}, which may refer to an individual word or a concept. Given document $d$, we first define the \emph{topic-specific verbosity} of document $d$, noted $v(t,d)$, which is the sum of frequencies of all words which belong to $t$:
\begin{equation}
v(t,d) = \sum_{ w \in {\mathcal{V}}} c(w,d) P(t|w)
\label{eq_appendix_topic_specific_verbosity}
\end{equation}
where $P(t|w)$ is the posterior probability that $w$ comes from $t$ (i.e., $\sum_{i=1}^M P(t_i|w) = 1$).

Under Eq. (\ref{eq_appendix_topic_specific_verbosity}), we readily show that $v(t,d)$ is the length of the passages in $d$ which belong to $t$.

By the definition above, we further show that the following equality holds:
\begin{eqnarray}
|d| & = & \sum_{i=1}^M v(t_i,d) \nonumber \\
    & = & v(t_1,d) + \cdots + v(t_M,d)
\label{eq_appendix_length_in_terms_of_topic_specific_verbosity}
\end{eqnarray}

To simplify Eq. (\ref{eq_appendix_length_in_terms_of_topic_specific_verbosity}), note that $v(t,d) = 0$ for most topics, as documents usually cover only a few topics.
Let $i_1 \cdots i_k \cdots i_{s(d)}$ be indexes of topics appearing in $d$ where $v(i_k,d) > 0$. Then, $|d|$ is reformulated as
\begin{eqnarray}
|d| & = & v(t_{i_1},d) + \cdots + v(t_ {i_{s(d)}},d)
\label{eq_appendix_length_in_terms_of_topic_specific_verbosity_derived}
\end{eqnarray}

Now, $v(d)$, the verbosity of document $d$, is defined as the average of all \emph{per-topic }verbosities computed for all $s(d)$ topics appearing in $d$, which is given by:
\begin{eqnarray}
v(d) & = & \frac{ v(t_{i_1},d) + \cdots + v(t_{i_{s(d)}},d) }{ s(d) } \nonumber \\
    & = & \frac{|d|}{ s(d) }
\label{eq_appendix_def_average_verbosity}
\end{eqnarray}

Thus, the average verbosity is the document length divided by the number of topics $s(d)$, which exactly replicates Eq. (\ref{eq_def_verbosity}). Given Eq. (\ref{eq_appendix_def_average_verbosity}), we only require $s(d)$, without need to estimate $v(t,d)$ which are usually unseen and hard to compute.

\section*{Appendix B: On Document-Specific Conjugate Prior for VN-DP}

The use of a document-specific conjugate prior for VN-DP (i.e., Eq. (\ref{eq_DirichletVN_prior})) is derived from the verbosity hypothesis. For convenience of discussion, suppose that $c'(w,d)$ indicates the unseen frequency of $w$ in $d$. Generally, smoothing uses $c_s(w,d)$ defined as $c(w,d) + c'(w,d)$ as a count of $w$ in $d$, thereby estimating $P(w|d)$ as $c_s(w,d)/\sum_{w \in V} c_s(w,d)$.

In DP, it is assumed that the \emph{pseudo length} $\mu$ is distributed over unseen words according to $P(w|C)$, resulting in $c'(w,d)$ = $\mu P(w|C)$.
In our case, however, because document length is decomposed to verbosity and scope, we need to introduce two pseudo factors for unseen words: verbosity and scope. Formally, let $v'(d)$ and $s'(d)$ be the verbosity and scope of an unseen part of $d$, respectively. Just like the formula of frequencies of seen words, which is given by $c(w,d)$ = $v(d) s(d) P_{ml}(w|d)$, frequencies of unseen words are formulated as $c'(w,d)$ = $v'(d) s'(d) P(w|C)$.

To determine $v'(d)$ and $s'(d)$, we use the following assumptions:

1. \emph{Verbosity of an unseen part: given a document, the verbosity of unseen passages (i.e., consisting of all unseen words) is the same as verbosity of the document.}
 -- The assumption is due to the verbosity hypothesis; $c(w,d)$ is mostly governed by $v(d)$. Just as the verbosity hypothesis is applied to seen words, we apply the verbosity hypothesis to unseen words. This results in $c'(w,d)$, the frequencies of unseen words, which should also be governed by $v(d)$.

2. \emph{Scope of an unseen part: given a document, the unseen scope of passages (i.e., consisting of all unseen words) is independent of the scope of the document.}
 -- Unlike verbosity, we do not make a document-specific setting for the scope, as the relation between the unseen scope of $d$ and $v(d)$ is not very clear.

Under these assumptions, we have $v'(d)$ = $v(d)$, and $s'(d)$ = $\mu$, thus resulting in $c'(w,d) = \mu v(d) P(w|C)$, which leads to our use of a document-specific prior in Eq. (\ref{eq_DirichletVN_prior}).

Therefore, the difference in formulating $c'(w,d)$ between DP and VN-DP results from whether or not we use the verbosity hypothesis for determining frequencies of unseen words.

\section*{Appendix C: Comparative Axiomatic Analysis}
In this appendix, we briefly summarize the derivations of $C_1$, $C_2$, and $C_3$ for Okapi and DP, where $C_1$ and $C_3$ are necessary but not sufficient for satisfying the particular constraint.  Let $d_1$ and $d_2$ be two given documents for LNCs and TF-LNC and $\Delta f(d_1,d_2,q)$ be $f(d_1,q) - f(d_2,q)$. All our derivations start from the inequality of $\Delta f(d_1,d_2,q) \geq 0$ (or $\Delta f(d1,d2,q) > 0$). For VN-DP, $\Delta f(d1,d2,q) \geq  0$ is equivalent to
\begin{equation}
\frac{s(d_1) p_{ml}(w|d_1) + \mu p(w|C)}{s(d_1) + \mu } \geq \frac{s(d_2) p_{ml}(w|d_1) + \mu p(w|C)}{s(d_2) + \mu }
\label{eq_appendix_score_difference_original_form_VN_DP}
\end{equation}

For VN-Okapi, $\Delta f(d_1,d_2,q) \geq 0$ is equivalent to
\begin{equation}
\frac{c(w,d_1)}{v(d_1)} \left( k \left( 1 - b + b \frac{s(d_1)}{avgs} \right) \right) idf(w)
\geq \frac{c(w,d_2)}{v(d_2)} \left( k \left( 1 - b + b \frac{s(d_2)}{avgs} \right) \right) idf(w)
\label{eq_appendix_score_difference_original_form_VN_Okapi}
\end{equation}

\subsection*{1)    LNC1}

We first show the derivation of the conditions for LNC1 under VN-DP and VN-Okapi. For the sake of convenience, we introduce the variables $m$ and $\varepsilon$ that are defined as $m$ = $v(d_1)/v(d_2)$ and $\varepsilon$ = $K/|d_1|$, respectively. According to the definition of PAN, we have the following relation between $s(d_1)$ and $s(d_2)$:
\begin{eqnarray}
s(d_2) & = & m (1 + \varepsilon) s(d_1) \nonumber \\
s(d_2) p_{ml}(w|d_2) & = & m \cdot s(d_1) p_{ml}(w|d_1)
\label{eq_appendix_LNC1_relation_s1_s2}
\end{eqnarray}
Below, we summarize the derivation for each case of VN-DP and VN-Okapi. \\

\subsubsection*{i)    VN-DP}

We first simplify Eq. (\ref{eq_appendix_score_difference_original_form_VN_DP}) by replacing $s(d_2)$ with the terms $m$, $\varepsilon$, and $s(d_1)$ using Eq. (\ref{eq_appendix_LNC1_relation_s1_s2}), as follows:
\begin{equation}
\mu \left( p_{ml}(w|d_1) - p(w|C) \right) \geq m \left( \mu \left( p_{ml}(w|d_1) - p(w|C) \right) - \mu \varepsilon p(w|C) - \varepsilon  s(d_1) p_{ml}(w|d_1) \right)
\label{eq_appendix_LNC1_equivalent_VN_DP_starting}
\end{equation}

First, when $p_{ml}(w|d) = p(w|C)$, it is easily shown that Eq. (\ref{eq_appendix_LNC1_equivalent_VN_DP_starting}) holds. Thus, we do not consider the equality case of $A_1$ to simplify Eq. (\ref{eq_appendix_LNC1_equivalent_VN_DP_starting}).

Under the remaining cases of $A_1$ (i.e., $p_{ml}(w|d) > p(w|C)$), Eq. (\ref{eq_appendix_LNC1_equivalent_VN_DP_starting}) is equivalent to
\begin{equation}
\frac{v(d_2)}{v(d_1)} \geq \left( 1 - \frac{K}{|d_1|} \frac{p(w|C) + p_{ml}(w|d_1) s(d_1) \mu^{-1} }{p_{ml}(w|d_1) - p(w|C) } \right)
\label{eq_appendix_LNC1_equivalent_VN_DP}
\end{equation}
Under $p_{ml}(w|d) > p(w|C)$, the right-hand side of Eq. (\ref{eq_appendix_LNC1_equivalent_VN_DP}) is a decreasing function with respect to $p(w|C)$, with the upper bound when $p(w|C)$ = 0. After replacing the right-hand side with this upper bound, the necessary condition for Eq. (\ref{eq_appendix_LNC1_equivalent_VN_DP}) is simplified to:
\begin{equation}
\frac{v(d_2)}{v(d_1)} \geq \left( 1 - \frac{K}{|d_1|} \frac{s(d_1)}{\mu}\right) = 1 - \frac{K}{v(d_1)} \frac{1}{\mu}
\label{eq_appendix_LNC1_necessary_VN_DP}
\end{equation}
which is satisfied if $C_1$ is true, regardless of the choice of parameter $\mu$.

\subsubsection*{ii)    VN-Okapi}

As in the case of VN-DP, we replace $s(d_1)$ with the terms $m$ and $\varepsilon$ based on Eq. (\ref{eq_appendix_LNC1_relation_s1_s2}), simplifying Eq. (\ref{eq_appendix_score_difference_original_form_VN_Okapi}) to
\begin{equation}
(1-b) idf(w) \geq m \left( (1-b) - b \varepsilon \frac{s(d_1)}{avgs} \right) idf(w)
\label{eq_appendix_LNC1_equivalent_VN_Okapi_starting}
\end{equation}

First, when $idf(w) = 0$, it is clear that Eq. (\ref{eq_appendix_LNC1_equivalent_VN_Okapi_starting}) holds.

Second, when $idf(w) > 0$ (under $A_1$), Eq. (\ref{eq_appendix_LNC1_equivalent_VN_Okapi_starting}) is further rewritten as
\begin{eqnarray}
\frac{1}{m} & \geq & 1 - \varepsilon \frac{b \cdot s(d_1)}{(1-b) avgs} \nonumber \\
& = &1 -  \frac{ b \cdot K}{(1-b) v(d_1) \cdot avgs}
\label{eq_appendix_LNC1_equivalent_VN_Okapi_derived}
\end{eqnarray}
which is equivalent to
\begin{eqnarray}
\frac{b \cdot K }{(1-b) avgs} \geq v(d_1) - v(d_2)
\label{eq_appendix_LNC1_equivalent_VN_Okapi}
\end{eqnarray}
Thus, LNC1 is satisfied if $C_1$ is true.

\subsection*{2)    LNC2}

We could straight forwardly derive that LNC2 is equivalent to $C_2$. To simplify the notation for the derivation, we introduce $\rho$ =$p_{ml}(w|d_1)$ = $p_{ml}(w|d_2)$; the equality holds because of the characteristic of PLS.
 We summarize the derivation of $C_2$ for each of VN-DP and VN-Okapi.\\

\subsubsection*{i)    VN-DP}

For VN-DP, Eq. (\ref{eq_appendix_score_difference_original_form_VN_DP}) is simplified to
\begin{equation}
\frac{s(d_1) \rho + \mu p(w|C)}{s(d_1) + \mu} \geq \frac{s(d_2) \rho + \mu p(w|C)}{s(d_2) + \mu}
\label{eq_appendix_LNC2_equivalent_VN_DP_starting}
\end{equation}
It is trivial to show that the necessary and sufficient condition for LNC2 is $C_2$, if $A_1$ holds:


\subsubsection*{ii)    VN-Okapi}

For VN-Okapi, Eq. (\ref{eq_appendix_score_difference_original_form_VN_Okapi}) is simplified to
\begin{equation}
\rho s(d_1) \left( k \left( 1 -b + b \frac{s(d_1) }{avgs}\right) \right) idf(w)
\geq \rho s(d_2) \left( k \left( 1 -b + b \frac{s(d_2) }{avgs}\right) \right) idf(w)
\label{eq_appendix_LNC2_equivalent_VN_Okapi_starting}
\end{equation}
Using $idf(w) \geq 0$ from $A_1$, Eq. (\ref{eq_appendix_LNC2_equivalent_VN_Okapi_starting}) is equivalent to $C_2$.\\

\subsection*{3)    TF-LNC}

For the sake of convenience, we introduce variables $m'$ and $\varepsilon'$ by putting $m'$ = $v(d_2)/v(d_1)$ and $\varepsilon'$ = $K/|d_2|$. According to the definition of PAR,

\begin{eqnarray}
s(d_1) & = & m' (1 + \varepsilon') s(d_2) \nonumber \\
s(d_1) p_{ml}(w|d_1) & = & m' \left( p_{ml}(w|d_2) + \varepsilon' \right) s(d_2)
\label{eq_appendix_TF_LNC_relation_s1_s2}
\end{eqnarray}
Below, we summarize the derivation for each of VN-DP and VN-Okapi. \\

\subsubsection*{i)    VN-DP}

We first simplify Eq. (\ref{eq_appendix_score_difference_original_form_VN_DP}) by replacing $s(d_1)$ with the terms $m'$, $\varepsilon'$, and $s(d_2)$, as follows:
\begin{eqnarray}
&& m' \left( \frac{\varepsilon' \left( 1 - p_{ml}(w|d_2) s(d_2)\right) }{\mu} + \left( p_{ml}(w|d_2) - p(w|C) \right) + \varepsilon' \left(1 - p(w|C) \right) \right)  \nonumber \\
&& >  \left( p_{ml}(w|d_2) - p(w|C) \right)
\label{eq_appendix_TF_LNC_equivalent_VN_DP_starting}
\end{eqnarray}

When $p_{ml}(w|d) =  p(w|C)$, it is easily shown that Eq. (\ref{eq_appendix_TF_LNC_equivalent_VN_DP_starting}) holds.

When $p_{ml}(w|d) >  p(w|C)$, in the remaining cases of $A_1$, Eq. (\ref{eq_appendix_TF_LNC_equivalent_VN_DP_starting}) is equivalent to
\begin{equation}
\frac{v(d_1)}{v(d_2)} < 1 + \frac{K}{|d_2|} \frac{\left(1-p(w|C) \right) + \left(1-p_{ml}(w|d_2) \right) s(d_2) \mu^{-1} }{ p_{ml}(w|d_2) - p(w|C)}
\label{eq_appendix_TF_LNC_equivalent_VN_DP_derived}
\end{equation}

For $p_{ml}(w|d) >  p(w|C)$, the right-hand side of Eq. (\ref{eq_appendix_TF_LNC_equivalent_VN_DP_derived}) is an increasing function with respect to $p(w|C)$, with the lower bound when $p(w|C)$ = 0. In addition, we can further lower the bound by eliminating  $(1-p(w|d_1))s(d_2)/\mu$ because it is a positive value. Thus, we obtain the following necessary condition for Eq. (\ref{eq_appendix_TF_LNC_equivalent_VN_DP_derived}):
\begin{equation}
\frac{v(d_1)}{v(d_2)} \leq 1  + \frac{K}{c(w,d_2)}
\label{eq_appendix_TF_LNC_equivalent_VN_DP}
\end{equation}
which is equivalent to $C_3$\footnote{Note that Eq. (\ref{eq_appendix_TF_LNC_equivalent_VN_DP}) can contain the equality condition, because $p(w|C) > 0$ (even when $p(w|C) \rightarrow 0$)}. \\

\subsubsection*{ii)    VN-Okapi}

As in the case of VN-DP, we replace $s(d_1)$ in the terms $m'$, $\varepsilon'$, and $s(d_2)$ using Eq. (\ref{eq_appendix_TF_LNC_relation_s1_s2}) simplifying Eq. (\ref{eq_appendix_score_difference_original_form_VN_Okapi}) to
\begin{eqnarray}
& & m' \left( \varepsilon' \left( 1 - p_{ml}(w|d)  \right) \frac{b \cdot s(d_2)}{avgs}
 + \left( p_{ml}(w|d) + \varepsilon' \right) (1-b) \right) idf(w) \nonumber \\
 & & \geq p_{ml}(w|d)(1-b) idf(w)
\label{eq_appendix_TF_LNC_equivalent_VN_Okapi_starting}
\end{eqnarray}

Under $A_1$, because $idf(w) \geq 0$, Eq. (\ref{eq_appendix_TF_LNC_equivalent_VN_Okapi_starting}) is equivalent to
\begin{eqnarray}
\frac{v(d_1)}{v(d_2)} < 1 + \frac{K}{|d_2|} \frac{(1-b) + \left( 1 - p_{ml}(w|d_2) \right) b \frac{s(d_2)}{avgs}}{p_{ml}(w|d_2)(1-b)}
\label{eq_appendix_TF_LNC_equivalent_VN_Okapi_derived}
\end{eqnarray}
A lower bound for the right-hand side Eq. (\ref{eq_appendix_TF_LNC_equivalent_VN_Okapi_derived}) is obtained by eliminating (1-$p_{ml}(w|d)) b s(d_2)/avgs$), which is a positive value. Therefore, the necessary condition for Eq. (\ref{eq_appendix_TF_LNC_equivalent_VN_Okapi_derived}) becomes
\begin{eqnarray}
\frac{v(d_1)}{v(d_2)} &\leq &1 + \frac{K}{|d_2|} \frac{1}{p_{ml}(w|d_2)} \nonumber \\
  & = & 1 + \frac{K}{c(w,d_2)}
\label{eq_appendix_TF_LNC_equivalent_VN_Okapi}
\end{eqnarray}
which is equivalent to $C_3$ \footnote{ Note that Eq. (\ref{eq_appendix_TF_LNC_equivalent_VN_Okapi}) can allow the equality condition.}.

\section*{Appendix D: Analysis Result of TF-LNC under UniqLength and LengthPower}
TF-LNC is true when using UniqLength and LengthPower. To prove this, let $\Delta_s$ be $s(d_1)-s(d_2)$. First, when $\Delta_s \geq K/v(d_2)$, it is equivalent to $v(d_1) \leq v(d_2)$, and thus, $C_3$ is satisfied. Otherwise (i.e., $\Delta_s < K/v(d_2)$), $C_3$ is equivalent to
\begin{equation}
c(w,d_2) \leq K \frac{v(d_2)}{v(d_1) - v(d_2)} = K \frac{|d_2| + \Delta_s \cdot v(d_2)}{K - \Delta_s \cdot v(d_2)}
\label{eq_C3_equivalent}
\end{equation}
The term on the right-hand side of Eq. (\ref{eq_C3_equivalent}) is a decreasing function with respect to $\Delta_s$. For the cases of UniqLength and LengthPower, $\Delta_s \geq 0$, regardless of $K$; therefore, Eq. (\ref{eq_C3_equivalent}) is satisfied if $c(w,d_2) \leq |d_2|$ (i.e., obtained by using $\Delta_s = 0$ in the right-hand side of Eq. (\ref{eq_C3_equivalent})), which is true for all cases:

\section*{Appendix E: Analysis Result of TF-LNC under EntropyPower}
In this appendix, for VN models using EntropyPower, we show that TF-LNC is satisfied if $C_4$ is true.
To prove this, let $\Delta_s$ be $s(d_1 )-s(d_2)$. When $\Delta_s \geq K/v(d_2 )$, it is equivalent to $v(d_1) \leq v(d_2)$, and thus, $C_3$ is satisfied. Otherwise, $C_3$ is equivalent to:.
\begin{eqnarray}
\Delta_s \geq \frac{K\left( c(w,d_2) - |d_2| \right)}{\left( c(w,d_2) + K \right) v(d_2)}
\label{eq_C3_equivalent_v2}
\end{eqnarray}

Eq. (\ref{eq_C3_equivalent_v2}) is further simplified to:
\begin{eqnarray}
\frac{s(d_1)}{s(d_2)} \geq \frac{c(w,d_2)}{c(w,d_2)+ 1} \frac{|d_2|+K}{|d_2|}
\label{eq_C3_equivalent_v2_simplified}
\end{eqnarray}

Using the definition of EntropyPower, we can rewrite $\log(s(d_1))$ and $\log(s(d_2))$ as:
\begin{eqnarray}
log\left( s(d_2) \right) = - \sum_{w' \in d_2} \frac{c(w',d_2)}{|d_2|} \log c(w',d_2) + \log |d_2|
\label{eq_log_scope_2_entropypower}
\end{eqnarray}

\begin{eqnarray}
\log\left( s(d_1) \right) = && -\frac{c(w,d_2) + K}{|d_2| + K} \log \left( c(w,d_2) + K \right) \nonumber \\
&&- \sum_{w' \in d_2, w' \neq w} \frac{c(w',d_2)}{|d_2|} \log c(w',d_2) + \log |d_2|
\label{eq_log_scope_1_entropypower}
\end{eqnarray}

Substituting Eqs. (\ref{eq_log_scope_1_entropypower}) and (\ref{eq_log_scope_2_entropypower}) in Eq. (\ref{eq_C3_equivalent_v2_simplified}), we now obtain the following condition for TF-LNC:
\begin{eqnarray}
\log\left( s(d_1) \right) - \log\left( s(d_2) \right) & = & -\frac{c(w,d_2) + K}{|d_2| + K} \log \left( c(w,d_2) + K \right) \nonumber \\
&& + \frac{|d_2|+K-1}{|d_2|\left( |d_2| + K \right)} \cdot c(w,d_2) \log\left( c(w,d_2) \right) \nonumber \\
&& + \frac{1}{|d_2| + K} \left( \log|d_2| - \log s(d_2) \right) \nonumber \\
&& + \log\left( |d_2| + K \right) - \log \left( |d_2| \right)
\label{eq_condition_TF_LNC_entropypower}
\end{eqnarray}


From the definition of TF-LNC, it is clear that once Eq. (\ref{eq_condition_TF_LNC_entropypower}) holds for $K$ = 1, then Eq. (\ref{eq_condition_TF_LNC_entropypower}) also holds for every $K$. Therefore, here, we only consider $K =1$. When $K = 1$, Eq. (\ref{eq_condition_TF_LNC_entropypower}) is further simplified to:
\begin{eqnarray}
\frac{|d_2| - c(w,d_2)}{|d_2| + 1} \log\left( c(w,d_2) + K \right) - && \frac{|d_2|+1-c(w,d_2)}{|d_2| + 1}\log c(w,d_2)  \nonumber \\
&& \geq \frac{1}{|d_2| + 1}\log \frac{s(d_2)}{|d_2|}
\label{eq_condition_TF_LNC_entropypower_simplified}
\end{eqnarray}
which leads to:

\begin{eqnarray}
\left( |d_2| - c(w,d_2)\right) \log \left( 1 + \frac{1}{c(w,d_2)} \right) - \log c(w,d_2) \geq \log \frac{s(d_2)}{|d_2|}
\label{eq_condition_TF_LNC_entropypower_simplified2}
\end{eqnarray}
Applying $(1+x)^n \geq (1+nx)$ to the first term of the left-hand side of Eq. (\ref{eq_condition_TF_LNC_entropypower_simplified2}), we obtain the following sufficient condition for Eq (\ref{eq_condition_TF_LNC_entropypower_simplified2}):
\begin{eqnarray}
\log\left( \frac{|d_2|}{c(w,d_2)} \right) - \log\left( c(w,d_2) \right) \geq \log \frac{s(d_2)}{|d_2|}
\label{eq_condition_TF_LNC_entropypower_simplified3}
\end{eqnarray}
which is rewritten as
\begin{eqnarray}
2\left( \log|d_2| - \log c(w,d_2) \right) \geq \log s(d_2)
\label{eq_condition_TF_LNC_entropypower_simplified4}
\end{eqnarray}

Therefore, we finally obtain the condition:
\begin{eqnarray}
s(d_2) \leq \left( \frac{|d_2|}{c(w,d_2)} \right)^2
\label{eq_condition_TF_LNC_entropypower_simplified4}
\end{eqnarray}
which is equivalent to $C_4$.


\end{document}